\documentclass[sigconf]{acmart} 
\usepackage{graphicx} 
\usepackage{subcaption}
\usepackage{siunitx}
\usepackage{listings}
\usepackage{xspace}
\usepackage[]{hyperref}
\usepackage{circledsteps}
\usepackage{float}
\usepackage{enumitem}
\usepackage{mathtools}
\usepackage{algorithm} 
\usepackage{algpseudocode} 
\usepackage[export]{adjustbox}
\usepackage{graphicx}
\usepackage{titlesec}

\usepackage{mdframed}
\mdfsetup{nobreak=true}
\usepackage{fancyhdr}

\titleformat*{\subsubsection}{\bfseries}

\usepackage[many,theorems,skins,breakable]{tcolorbox}
\definecolor{tcbcolback}{RGB}{245,243,253}
\definecolor{tcbcolbox}{RGB}{254 218 173}
\definecolor{tcbcolframe}{RGB}{245,243,253}
\newtcolorbox{visionbox}[2][]{%
    colback=white!12,
    coltitle=black,
    colframe=teal!50,
    fonttitle=\bfseries,
    title=#2, 
    sharp corners,
    rounded corners=southeast,
    boxrule=0pt,
    enhanced,
    drop fuzzy shadow,
    #1, 
    top=3pt,bottom=2pt,left=3pt,right=3pt
    }

\AtBeginDocument{%
  }

\copyrightyear{2025} 
\acmYear{2025} 
\setcopyright{cc}
\setcctype{by}
\acmConference[HPDC '25]{The 34th International Symposium on High-Performance Parallel and Distributed Computing}{July 20--23, 2025}{Notre Dame, IN, USA}
\acmBooktitle{The 34th International Symposium on High-Performance Parallel and Distributed Computing (HPDC '25), July 20--23, 2025, Notre Dame, IN, USA}
\acmDOI{10.1145/3731545.3731576}
\acmISBN{979-8-4007-1869-4/2025/07}



\title{CarbonEdge: Leveraging Mesoscale Spatial Carbon-Intensity Variations for Low Carbon Edge Computing}

\renewcommand{\shorttitle}{CarbonEdge}
\author{Li Wu}
\affiliation{
  \institution{University of Massachusetts Amherst}
  \country{USA}
}
\author{Walid A. Hanafy}
\affiliation{
  \institution{University of Massachusetts Amherst}
  \country{USA}
}
\author{Abel Souza}
\affiliation{
  \institution{University of California, Santa Cruz}
  \country{USA}
}
\author{Khai Nguyen}
\affiliation{
  \institution{University of Massachusetts Amherst}
  \country{USA}
}
\author{Jan Harkes}
\affiliation{
  \institution{Carnegie Mellon University}
  \country{USA}
}
\author{David Irwin}
\affiliation{
  \institution{University of Massachusetts Amherst}
  \country{USA}
}
\author{Mahadev Satyanarayanan}
\affiliation{
  \institution{Carnegie Mellon University}
  \country{USA}
}
\author{Prashant Shenoy}
\affiliation{
  \institution{University of Massachusetts Amherst}
  \country{USA}
}

\def\carbonunit{g$\cdot$CO$_2$eq/kWh\xspace}
\def\emissionunit{g$\cdot$CO$_2$eq\xspace}

\def\proposedsystem{\textit{CarbonEdge}\xspace}

\def\latencyaware{\texttt{Latency-aware}\xspace}
\def\energyaware{\texttt{Energy-aware}\xspace}
\def\intensityaware{\texttt{Intensity-aware}\xspace}

\titleformat{\section}{\Large\bfseries\MakeUppercase}{\thesection}{1em}{}
\begin{document}

\begin{abstract}

The proliferation of latency-critical and compute-intensive edge applications is driving increases in computing demand and carbon emissions at the edge. To better understand carbon emissions at the edge, we analyze granular carbon intensity traces at intermediate "mesoscales," such as within a single US state or among neighboring countries in Europe, and observe significant variations in carbon intensity at these spatial scales. Importantly, our analysis shows that carbon intensity variations, which are known to occur at large continental scales (e.g., cloud regions), also occur at much finer spatial scales, making it feasible to exploit geographic workload shifting in the edge computing context.
Motivated by these findings, we propose \proposedsystem, a carbon-aware framework for edge computing that optimizes the placement of edge workloads across mesoscale edge data centers to reduce carbon emissions while meeting latency SLOs. We implement \proposedsystem and evaluate it on a real edge computing testbed and through large-scale simulations for multiple edge workloads and settings. Our experimental results on a real testbed demonstrate that \proposedsystem can reduce emissions by up to 78.7\% for a regional edge deployment in central Europe. Moreover, our CDN-scale experiments show potential savings of 49.5\% and 67.8\% in the US and Europe, respectively, while limiting the one-way latency increase to less than 5.5 ms.

\end{abstract}


\maketitle

\fancyhead[R]{Wu et al.}

\section{Introduction}
\label{sec:introduction}
Data centers consumed more than 460 terawatt-hours (TWh) of energy in 2022, and are expected to consume more than 1000 TWh by 2026~\cite{iea2024electricity}. As a result, data centers are already generating roughly 1\% of global carbon emissions and could emit more than 2.5 billion metric tons of $\text{CO}_2$ by the end of the decade. The sustainable growth of data center capacity has emerged as a critical challenge in our society's transition to a low-carbon future, especially with the accelerating build-out of data center capacity to satisfy the growing demand for AI workloads. Historically, cloud operators have addressed data center sustainability issues by optimizing their energy efficiency (i.e., their computational work done per unit of energy consumed).  However, optimizing energy-efficiency alone will likely not be sufficient to satisfy cloud platforms' carbon emissions targets~\cite{Bashir2022:HotAir}.  In particular, since data center energy-efficiency is already highly optimized after years of research, further optimizations are expected to yield diminishing marginal improvements moving forward. 


As a result, researchers have recently focused on several alternative approaches for reducing carbon footprint of cloud data centers and improving their carbon efficiency (i.e., their computational work done per unit of carbon emitted). Specifically, to optimize hyperscale data centers' carbon efficiency, cloud operators have deployed both supply- and demand-side approaches.  On the supply-side, cloud operators have procured green energy (e.g., wind or solar) through long-term contracts to power their data center operations \cite{google_offshore_wind_2024}, while on the demand-side, researchers have explored techniques for modulating data center workload demand and its resulting carbon emissions to optimize their carbon footprint \cite{wait-awhile}. Since the electric grid in different regions uses different mixes of generation sources, grids with a higher penetration of low-carbon sources, such as hydro, solar, or wind, tend to produce lower-carbon electricity. Spatial workload shifting approaches exploit these regional differences in energy's carbon intensity by proactively shifting workloads to data center locations with lower-carbon energy, thereby performing the same computation while incurring fewer emissions. Recent research has shown that cloud workloads, such as machine learning training and batch processing, are amenable to such spatial shifting optimizations and can yield significant reductions in applications' carbon footprint~\cite{sukprasert2024limitations, cloudcarbon, Gsteiger2024:Caribou, Murillo2024:CDNShifter}. However, these optimizations typically incur large network delays to migrate workloads over long distances to a different data centers, and thus can increase user latency for interactive workloads that are latency-sensitive.

Since spatial differences in the grid's carbon intensity are clearly evident over large geographical distances, spatial shifting has largely been studied for cloud workloads at continental scales, i.e., across entire continents or between continents.  For example, shifting workloads from eastern to western North America, or shifting workloads from North America or Asia to Europe. The differences in grid carbon intensity at these scales are due to the vastly different generation mixes at distant locations. { \em As a result, conventional wisdom has held that spatial workload shifting is unsuitable for edge data centers, since moving edge workloads over such long distances to distant edge data centers would result in unacceptable increases in the latency of edge applications.} Consequently, edge data centers thus far have not leveraged this key carbon optimization technique. 

In this paper, we challenge this conventional wisdom and show that spatial workload shifting is a feasible carbon-optimization approach for edge data centers deployed in many, although not all, parts of the world. Our key insight is that spatial differences in grid carbon intensity do frequently occur even at ``mesoscales'' (i.e., smaller distances of tens to a few hundred kilometers), especially as the penetration of wind and solar renewables continues to grow.  While, on average, variations in carbon intensity are certainly larger at longer distances, there are meaningful differences in energy's carbon intensity at short distances in many parts of the world.  Such mesoscale differences open up new opportunities for spatial workload shifting across nearby edge data centers, enabling edge workloads to optimize their carbon footprint with limited performance impact on latency-sensitive applications. In contrast, as mentioned above, prior work has focused primarily on exploiting spatial differences in energy's carbon intensity at large continental scales, i.e., across a thousand kilometers or more, where variations in energy's carbon intensity arise from large environmental differences.  For example, at continental scales, it may be daytime in one location with plentiful solar generation and nighttime in another with zero solar generation. Instead, mesoscale differences in carbon intensity generally arise from differences in a location's specific mix of various generation sources, e.g., hydro, coal, natural gas, oil, solar, wind, nuclear, etc., and types of generators.  For example, a municipal utility that serves a small town may have its own low-carbon hydro-generating plant, while nearby towns are served by a private utility that generates most of its power from high-carbon fossil fuels. Such differences give rise to variations in energy's carbon intensity, even at relatively short distances. 



Motivated by these observations, this paper presents \proposedsystem, a carbon-aware orchestration framework for distributed edge data centers that supports spatial workload shifting at mesoscales. \proposedsystem optimizes workload placement to significantly reduce carbon emissions of edge applications within a mesoscale region while satisfying latency constraints. 
Importantly, \proposedsystem considers the diversity of edge applications and resource heterogeneity when determining workload placement, which affects how applications consume energy at specific locations. This aspect is crucial because the carbon emissions of applications depend on both their energy consumption and the carbon intensity of the energy used. We hypothesize that 
small spatial-scale variations in carbon intensity can enable 
\proposedsystem to reduce the operational carbon footprint of edge applications without significantly impacting their low-latency benefits.
In designing, implementing, and evaluating \proposedsystem, we make the following contributions. 

\begin{enumerate}[leftmargin=*]
    \item \textbf{Mesoscale Carbon Analysis.} Our analysis is the first to demonstrate significant variations in grid carbon intensity at mesoscale distances, thereby making it feasible to deploy workload shifting optimizations in edge computing platforms.
    We present a detailed empirical analysis of the granular carbon intensity data and latency traces of 148 regions in the world (\autoref{sec:carbon_analysis}).

    

   \item \textbf{\proposedsystem Design and Implementation.} Based on our findings, we propose \proposedsystem, a carbon-aware placement framework to reduce carbon emissions from edge data centers at mesoscales. \proposedsystem integrates carbon intensity variations across edge data center locations and accounts for energy-efficiency differences among heterogeneous resources to intelligently distribute edge workloads to minimize carbon emissions (\autoref{sec:design}). Additionally, we implement a full prototype of a carbon-aware edge orchestration framework on top of Sinfonia, a Kubernetes-based framework for edge data centers, and plan to release it as open source (\autoref{sec:implementation}). 
   
   
    \item \textbf{Experimental Evaluation.} We evaluate \proposedsystem in both edge testbeds and large-scale simulations, using real-world traces, edge workloads, and diverse edge settings. Our experimental results on real testbed demonstrate that \proposedsystem can reduce emissions by up to 78.7\% in mesoscale regional edge deployments. Furthermore, our CDN-scale simulations indicate that \proposedsystem yields 49.5\% and 67.8\% savings in the US and Europe, respectively, while limiting the one-way latency increase to less than 5.5 ms 
    (\autoref{sec:evaluation}).
    
    
\end{enumerate}

\section{Background}
\label{sec:background}
This section provides background on grid energy's carbon intensity, carbon-aware workload optimization, and edge data centers. 

\subsection{Electric Grid Carbon Intensity}

\begin{figure}[t]
    \centering
    \subfloat[\centering Energy Mix]{{
        \includegraphics[width=0.45\linewidth]{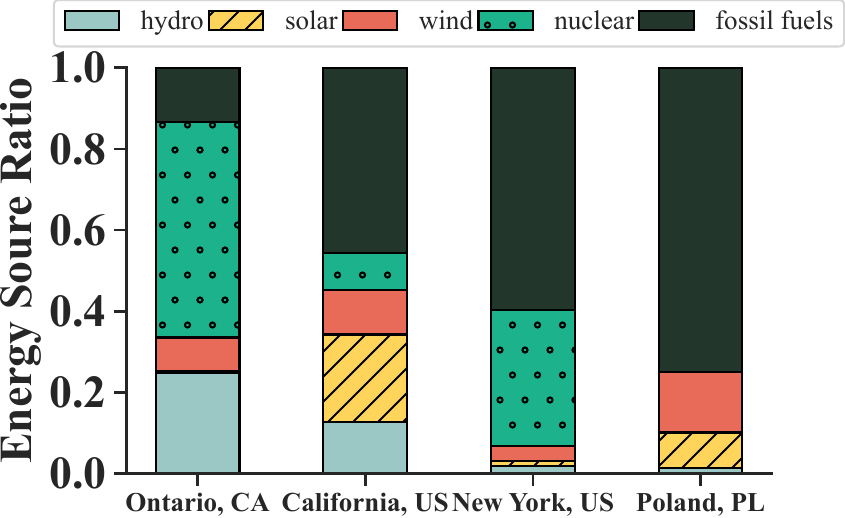}}
        \label{fig:cloud_energy_source}
    }%
    \quad
    \subfloat[\centering Carbon Intensity ]{{
        \includegraphics[width=0.45\linewidth]{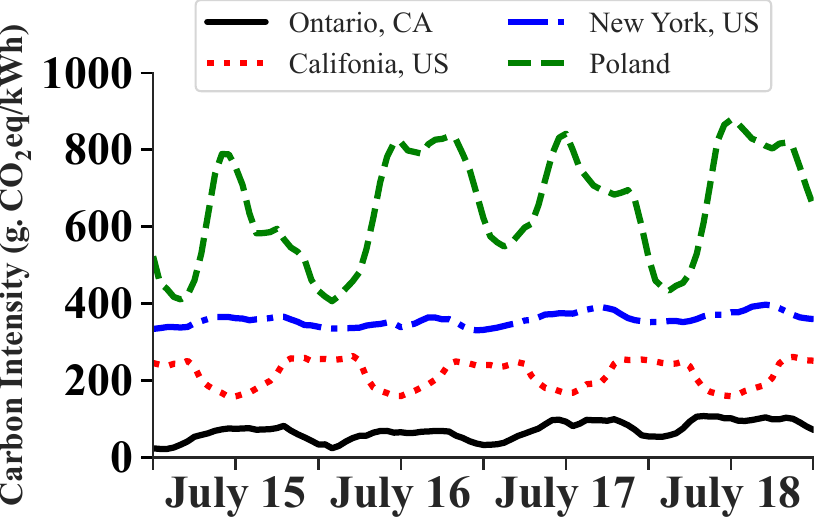}}
        \label{fig:cloud_ci}
    }%
    \caption{Energy mix and carbon intensity of four regions.}
    \label{fig:cloud_example}
    \vspace{-6mm}
\end{figure}

The electricity supplied by the electric grid at a given location comes from a mix of generation sources, such as natural gas, coal, hydro, solar, and wind. The relative proportion of generation from each source varies from one region to another, depending on the types of generation sources present in each region. For example, as shown in ~\autoref{fig:cloud_energy_source}, in the Ontario region of Canada, most energy comes from nuclear and hydroelectric energy sources. At the same time, eastern European countries such as Poland have a more significant proportion of coal and natural gas. The carbon intensity of electricity is defined as the total $\text{CO}_2$ emissions per unit of electricity generation and is measured in \carbonunit. For each location, it is computed as the weighted average of the carbon intensity of the source energy mix at that location. \autoref{fig:cloud_ci} shows the carbon intensity of the energy supply in four different countries and regions ---  Ontario region of Canada, California and New York in the US, and Poland in Europe --- and shows that there are significant differences in the carbon intensity of electricity at the spatial granularity of countries or large geographic regions. 

As grid operators have begun to report their real-time energy generation mixes, third-party carbon information services, such as Electricity Maps~\cite{electricity-map} and WattTime~\cite{watttime}, have begun exposing this carbon intensity data to data center operators and applications via real-time APIs and forecasting services. Our paper assumes the availability of such carbon intensity data for carbon optimizations in edge data centers.

\begin{figure*}[t]
    \centering%
    \hfill
    \begin{subfigure}{0.225\linewidth}%
        \centering
        \includegraphics[width=\linewidth]{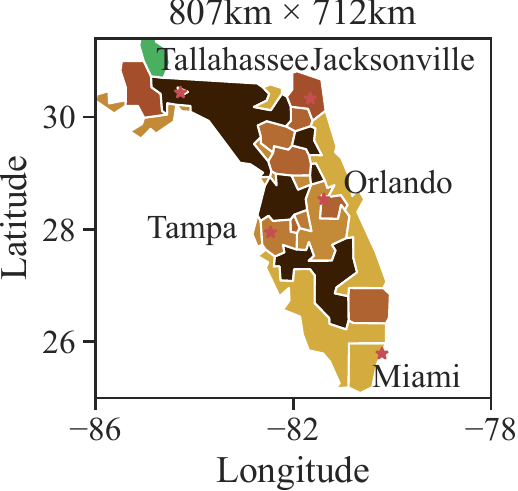}%
        \caption{Florida}
        \label{fig:cv_snapshot_r1}
    \end{subfigure}%
    \hfill
    \begin{subfigure}{0.21\linewidth}%
        \centering
        \includegraphics[width=\linewidth]{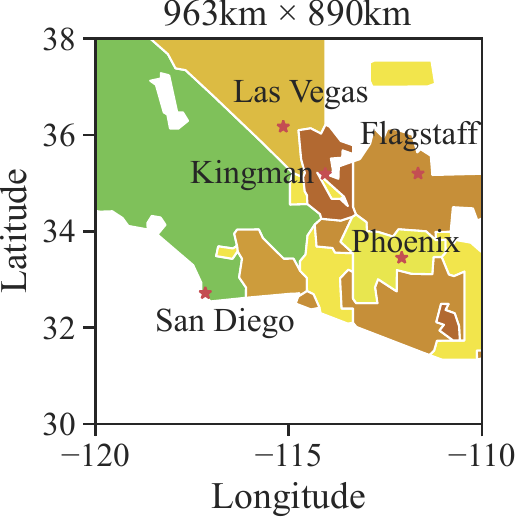}%
        \caption{West US}
        \label{fig:cv_snapshot_r2}
    \end{subfigure}%
    \hfill
    \begin{subfigure}{0.19\linewidth}%
        \centering
        \includegraphics[width=\linewidth]{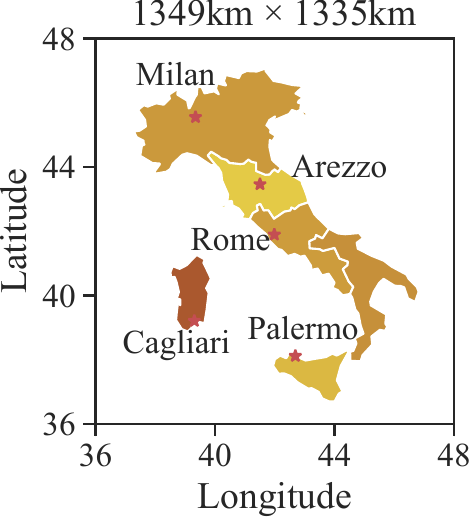}%
        \caption{Italy}
        \label{fig:cv_snapshot_r3}
    \end{subfigure}%
    \hfill
    \begin{subfigure}{0.175\linewidth}%
        \centering
        \includegraphics[width=\linewidth]{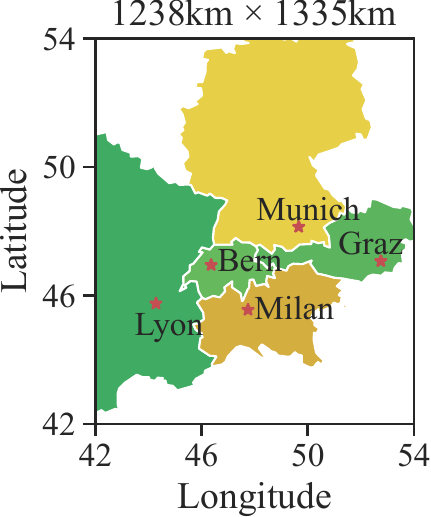}%
        \caption{Central EU}
        \label{fig:cv_snapshot_r4}
    \end{subfigure}%
    \includegraphics[width=0.085\linewidth]{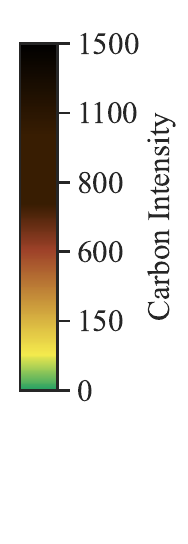}
    \hfill
    \hfill 
    \caption{Carbon intensity snapshots of four mesoscale regions, highlighting variations across zones.}%
    \label{fig:cv_snapshot}%
\end{figure*}

\subsection{Carbon-aware Workload Optimizations}

The availability of real-time carbon intensity data has motivated cloud providers and applications to schedule workloads based on variations in the carbon intensity of electricity. The resulting carbon-aware scheduling approaches can be broadly viewed as workload shifting, which exploits temporal and spatial variations in the carbon intensity of electricity. Temporal workload shifting exploits temporal fluctuations in carbon intensity at a given location by scheduling (or delaying) jobs to periods of low carbon intensity. Such techniques are well-suited for batch workloads, which have temporal flexibility and can tolerate delays in their completion times.  There have been numerous recent works that leverage temporal workload shifting to reduce the carbon emissions of batch workloads in the cloud~\cite{acun2023carbon, wait-awhile, sukprasert2024limitations, ecovisor, hanafy2023carbonscaler}.  

In contrast, spatial workload shifting exploits spatial variations in carbon intensity across locations by moving jobs or requests to data center locations with a lower carbon intensity supply. 
Such spatial shifting optimizations have been studied in the cloud context by moving cloud workloads across data center locations that span large geographic regions, countries, or even continents \cite{cloudcarbon,sukprasert2024limitations, Gao-2012-being-green, Gsteiger2024:Caribou, Murillo2024:CDNShifter}. 
For example, as depicted in \autoref{fig:cloud_ci}, cloud workloads can be moved from the New York region of a public cloud to the Ontario region, whose electricity supply has a lower carbon intensity. 
In theory, spatial shifting can be implemented for both interactive workloads, such as web services, as well as batch workloads. In practice, however, migrating interactive requests to distant data centers increases their user-perceived latency, and hence, spatial shifting has been primarily utilized for batch applications, such as machine learning training~\cite{cloudcarbon}.  Prior work has also shown that spatial shifting generally has much more potential for reducing carbon compared to temporal shifting~\cite{sukprasert2024limitations}.  This insight derives from the fact that there tend to be much larger differences in carbon between locations than within any one location over time.


\subsection{Edge Data Centers}

Edge computing involves deploying computing resources in the form of small server clusters at the network's edge close to end users. Edge computing is well-suited for low latency services since it avoids network delays incurred by traversing to more distant cloud data centers. For example, 
regional edge clusters, deployed by edge or even cloud providers, have been used to host latency-sensitive applications such as mobile offloading, augmented reality (AR), and deep learning inference \cite{Satya17_emergence,Satya09_Cloudlets}. Another example is a content delivery network that operates large geo-distributed edge clusters and serves web and multimedia content to users from proximate edge locations.



While edge data centers can optimize their carbon footprint via temporal workload shifting, such methods are ill-suited for interactive or latency-sensitive applications that are prevalent at the edge, since such workloads cannot be delayed or time-shifted. In contrast, spatial shifting optimizations have traditionally been performed at larger continental or global scales, i.e., across entire continents or across multiple continents, to exploit carbon intensity variations present at that scale~\cite{cloudcarbon, Gsteiger2024:Caribou, sukprasert2024limitations}. While these methods work well for cloud-based batch workloads, spatial shifting of interactive edge workloads at such large scales results in large latency increases. Hence, spatial shifting has not been considered for edge applications in prior work. We argue that spatial shifting is feasible even in the edge context by exploiting mesoscale variations in grid carbon intensity that are beginning to appear in today's energy grids.

\section{Mesoscale Carbon Analysis}
\label{sec:carbon_analysis}

This section presents an empirical study of grid carbon intensity differences that occur over mesoscale geographic distances of tens to hundreds of kilometers. We also analyze the increases in network latency at these scales. Our empirical study seeks to answer two key questions. 

\begin{enumerate}[leftmargin=*]
    \item {\em How much does energy's carbon intensity vary within mesoscale regions that span tens to hundreds of kilometers, and are these differences large enough to warrant the use of spatial workload shifting in distributed edge data centers?}
    
    \item {\em How prevalent are these types of mesoscale variations in different parts of the world? Are they sufficiently common to warrant the broad deployment of carbon optimization techniques in edge data centers across the world?}
\end{enumerate}


\begin{figure}[t]
  \centering%
 \begin{subfigure}{0.45\linewidth}%
 \centering
       \includegraphics[width=\linewidth]{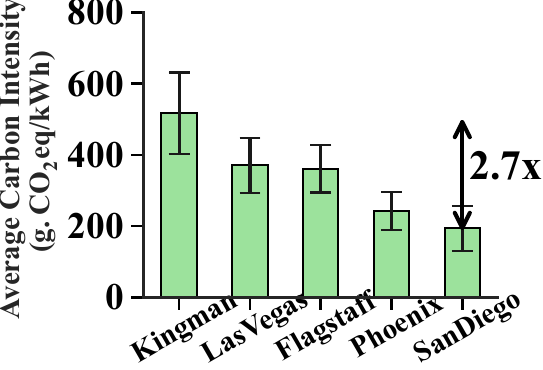}%
       \caption{West US}
       \label{fig:cv_yearly_r2}
    \end{subfigure}%
\hfill%
\begin{subfigure}{0.45\linewidth}%
        \centering
       \includegraphics[width=\linewidth]{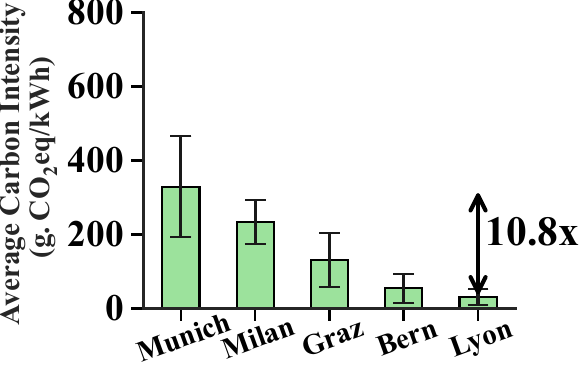}%
       \caption{Central EU}
       \label{fig:cv_yearly_r3}
    \end{subfigure}%
    \caption{Yearly carbon intensity of two mesoscale regions.}
    \label{fig:cv_yearly}%
\end{figure}

\begin{figure}[t]
    \centering%
    \includegraphics[width=0.9\linewidth]{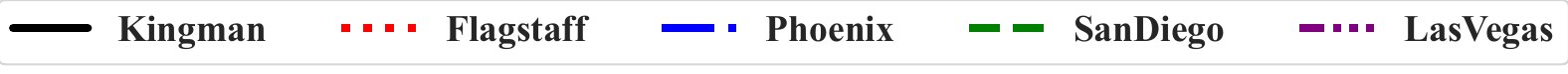}\\
    \begin{subfigure}{0.45\linewidth}%
    \centering%
    \captionsetup{justification=centering}
    \includegraphics[width=\linewidth]{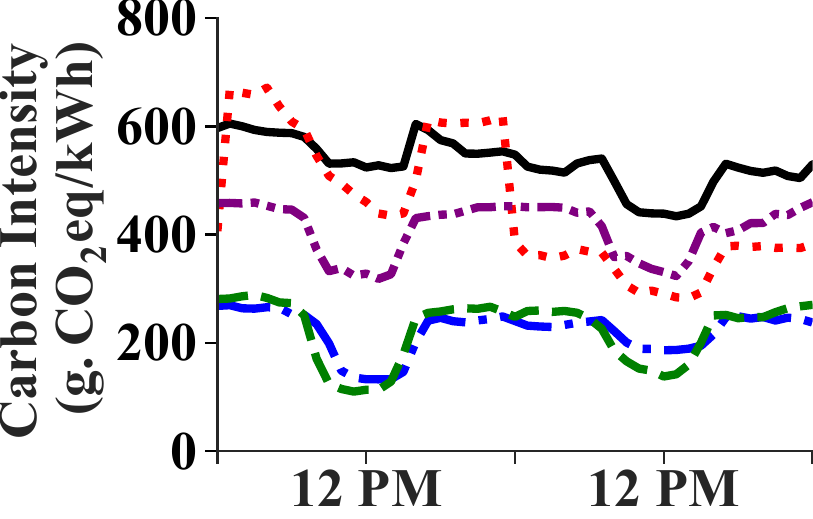}
    \caption{Two-day (Dec 25-27)}
    \label{fig:carbon_intensity_temporal_days_WUS}
    \end{subfigure}
    \quad
    \begin{subfigure}{0.45\linewidth}%
    \centering%
    \captionsetup{justification=centering}
    \includegraphics[width=\linewidth]{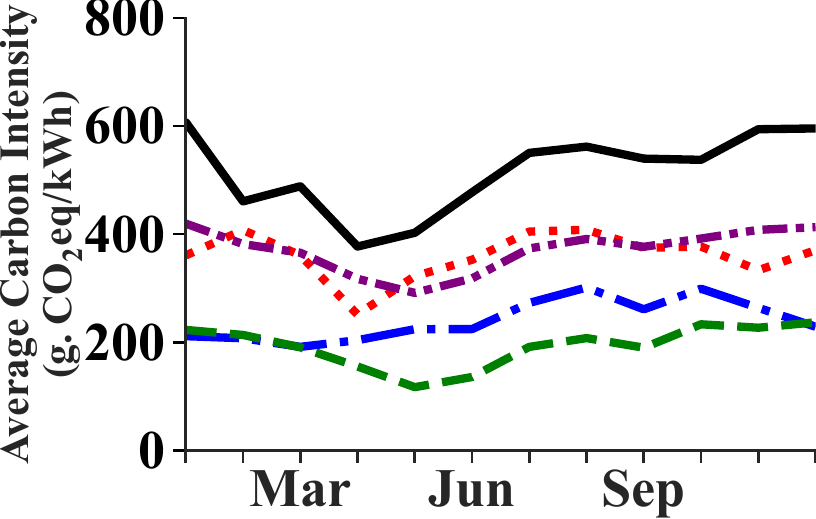}
    \caption{Year-long (2023)} 
    \label{fig:carbon_intensity_temporal_months_WUS}
    \end{subfigure}
    \caption{Spatial-temporal variations in carbon intensity over two days and 12 months in 2023 in the West US.}
    \label{fig:carbon_intensity_temporal}
    \vspace{-5mm}
\end{figure}

\begin{figure*}[tb]
    \centering
    \subfloat[\centering D = 200 km ]{\includegraphics[width=0.22\linewidth]{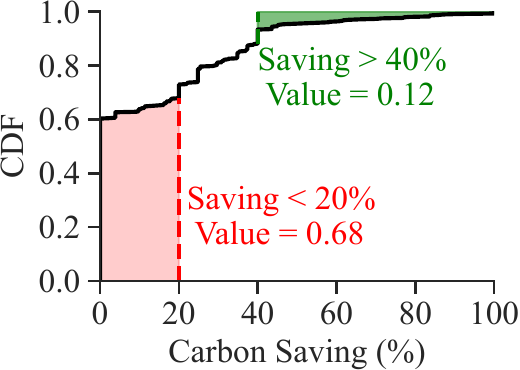}%
    \label{fig:carbon_saving_200km}%
    }%
    \hfill
    \subfloat[\centering D = 500 km ]{\includegraphics[width=0.22\linewidth]{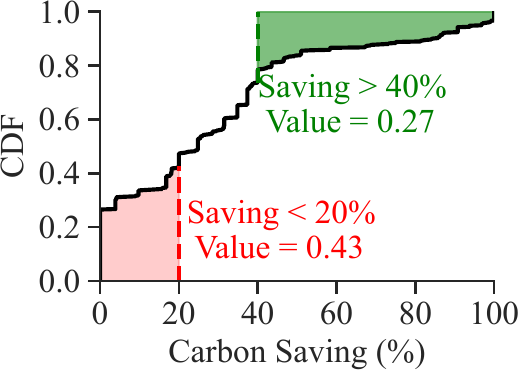}%
    \label{fig:carbon_saving_500km}%
    }%
    \hfill
    \subfloat[\centering D = 1000 km ]{\includegraphics[width=0.22\linewidth]{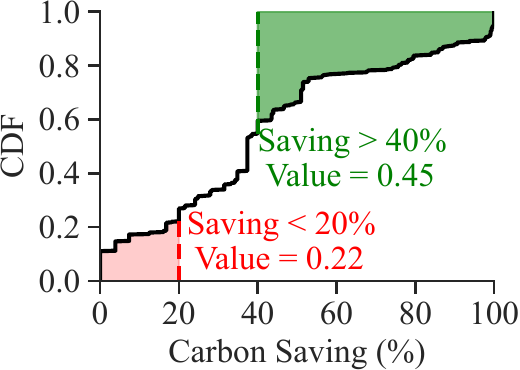}
    \label{fig:carbon_saving_1000km}
    }%
    \hfill
    \subfloat[\centering Radius-Latency ]{\includegraphics[width=0.22\linewidth]{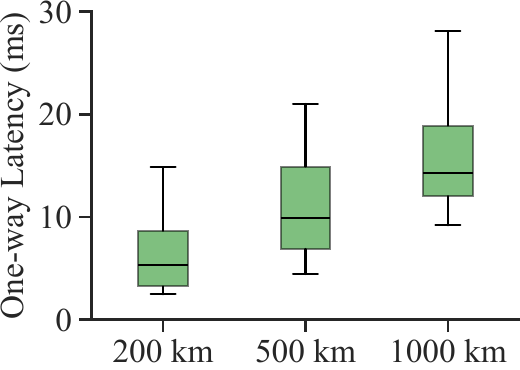}%
    \label{fig:mesoscale_round_trip_latency}%
    }%
    \caption{Carbon savings with search radii of 200 km, 500 km, and 1000 km. (d) One-way latency across pairwise distances.}
    \label{fig:carbon_saving_distance}
\end{figure*}

\subsection{Carbon Intensity Analysis at Mesoscales}

To understand the differences in grid carbon intensity that are seen at mesoscales, we conducted a measurement study where we collected carbon intensity traces for 148 carbon zones worldwide for an entire year (2023). For the purpose of our study, a carbon zone, or simply a \textit{zone}, is a geographic area whose grid operator provides carbon intensity data. The geographic size of a carbon zone depends on the area served by the grid operator and can vary from a city to an entire state or even a small country. 
Further, we also collected round-trip latency traces from the WonderNetwork~\cite{wonder-proxy-2020}, which provides ping traces (in milliseconds) to cities across the world. We describe our data sources in \autoref{sec:real_world_traces}. 

To illustrate the carbon intensity differences at the mesoscale, we first select four specific mesoscale regions, each comprising five carbon zones, across the United States and Europe. \autoref{fig:cv_snapshot} depicts a heat-map of the carbon intensity variations within each mesoscale region for a single hour in 2023, with darker colors representing higher carbon intensity values. We assume that each of the five carbon zones within a mesoscale region has an edge data center. 
The Florida region, for example, consists of five cities, each hosting an edge data center, that is a few hundred kilometers apart from one another. 
The figure shows significant differences in carbon intensity values even at this scale, with inter-zone variations of 2.5$\times$ in Florida, 7.9$\times$ in the west US, 2.2$\times$ in Italy, and 19.5$\times$ in Central Europe. 

~\autoref{fig:cv_yearly} then plots the mean carbon intensity over the entire year for two regions. The figure confirms that the differences in carbon intensity persist across the year. Furthermore, the average difference between the maximum and minimum carbon intensity across zones in a region is 
2.7$\times$ in the west US, 
and 10.8$\times$ in central Europe. Importantly, these differences compare favorably to those reported across global cloud regions. For instance, a recent study of spatial differences in carbon intensity across Amazon's cloud regions reported an order of magnitude difference across AWS cloud regions in Europe and Asia~\cite{sukprasert2024limitations}.
Moreover, since the relative mix of energy sources changes over time, \autoref{fig:carbon_intensity_temporal} shows temporal fluctuations in the carbon intensity of edge data centers within each mesoscale region within a day (\autoref{fig:carbon_intensity_temporal_days_WUS}) 
and across seasons (\autoref{fig:carbon_intensity_temporal_months_WUS}). 
For instance, Flagstaff, AZ (see \autoref{fig:carbon_intensity_temporal_days_WUS}) exhibits a daily difference of $\sim$300\carbonunit.
~\autoref{fig:carbon_intensity_temporal_months_WUS} shows how monthly average carbon intensity changes.
For example, Kingman, AZ, exhibits a $\sim$200 \carbonunit change between March and November due to its reliance on solar energy. 
\begin{table}[t]
  \centering
    \caption{One-way network latency (ms).  
    } 
  \label{tab:latency}
  \subfloat[Florida]{%
  \resizebox{0.53\linewidth}{!}{
  \begin{tabular}{|c|c|c|c|c|}
    \hline
    Location &  Miami &  Orlando &  Tampa & Tallah. \\
    \hline
    Jacksonville & 3.64 & 5.32 & 6.86 & 3.42\\ \hline
    Miami &   - & 4.5  & 3.37 & 7.2 \\ \hline
    Orlando &   & - &  1.86 & 4.35\\ \hline
    Tampa &   &  & - & 4.14\\ \hline
    Tallahassee  &  & &  & -\\ \hline
  \end{tabular}
  \label{tab:latency_r1} 
  }
  }
  \subfloat[Central EU]{%
  \resizebox{0.47\linewidth}{!}{
    \begin{tabular}{|c|c|c|c|c|c|}
    \hline
    Location &  Graz &  Lyon &  Milan &  Munich  \\
    \hline
    Bern, CH & 8.78 & 6.28 & 6.45 &3.985\\ \hline
    Graz, AT  & - & 16.22  & 11.98 & 8.36\\ \hline
    Lyon, FR  &  & - &  9.34 & 8.82\\ \hline
    Milan, IT   &  &  & - & 8.65\\ 
    \hline
    Munich, DE   &  &  &  & -\\ 
    \hline
  \end{tabular}
  }
  } 
  \vspace{-3mm}
\end{table}

Finally, ~\autoref{tab:latency} shows the pairwise one-way network latency between edge data centers, within two mesoscale regions. The table shows that, unsurprisingly, the latency grows with geographic distance. However, the increase in latency due to shifting workload from one edge location to another ranges from a few milliseconds to $\sim$16 ms, depending on the distance and the network topology between locations.

\noindent \textit{\textbf{Key Takeaways.} Our results show significant differences in the carbon intensity of electricity at mesoscale distances, similar to those reported at continental scales between cloud regions. These mesoscale variations demonstrate the feasibility of using spatial workload-shifting optimizations for edge data centers.}

\subsection{Mesoscale Analysis across Continents}
Having shown that there can be significant differences in carbon intensity at the mesoscale, a key question is whether such differences are commonplace in different parts of the world or confined to a few specific locations. To answer this question, we conduct an analysis of carbon intensity traces across 496 Akamai edge data centers in the United States and Europe. For each edge data center, we find the location with the highest carbon intensity difference within a threshold radius distance $D$ and compute the percentage difference in carbon intensity between the two locations. ~\autoref{fig:carbon_saving_distance} plots a CDF of the observed pairwise differences for different values of threshold radius $D$ (from $D$ = 200 km to $D$ = 1000 km).

For a radius of 200 km, ~\autoref{fig:carbon_saving_200km} shows that 32\% of the edge data centers have at least one data center with a carbon intensity difference of more than 20\%, and 12\% of locations have a data center with a carbon intensity difference of more than 40\%. At the same time, 68\% of the edge data centers do not have any location with a significant spatial carbon intensity difference (i.e., more than 20\%). As the radius increases, the chances of finding an edge location with significant carbon intensity differences grow.
As shown in \autoref{fig:carbon_saving_500km} and \autoref{fig:carbon_saving_1000km}, increasing the radius to 500 km and 1000 km allows 57\% and 78\% of edge data centers to reduce their emissions by more than 20\%. In addition, this increase enables 27\% and 45\% of edge data centers to {\em significantly} reduce their carbon emissions by more than 40\% for the 500 km and 1000 km radius, respectively. The fraction of edge locations without any significant carbon intensity differences within its radius falls to 22\% for $D=1000$ km. Lastly, \autoref{fig:mesoscale_round_trip_latency} shows that the median increase in latency ranges from 5.3 ms for $D=200$ km to 14.3 ms for $D=1000$ km.

\noindent \textit{\textbf{Key Takeaways.} More than 78\% of the edge locations in Europe and North America see carbon intensity differences exceeding 20\% within a radius of 1000 km, indicating that mesoscale carbon intensity variations are prevalent in many regions of the world.\footnote{Our analysis could not be extended to other continents (e.g., Asia, Australia) due to the unavailability of fine-grain spatial carbon intensity data, but we anticipate similar trends will persist as the adoption of renewables continues to grow globally.}}

\section{\proposedsystem Design and Policies }
\label{sec:design}
Motivated by our findings from the mesoscale carbon analysis, we introduce a carbon-aware framework for edge computing, named \proposedsystem, which employs the variations of carbon intensity across edge data centers to intelligently distribute edge applications while satisfying the low-latency demands. We formalize our carbon-aware edge placement problem with latency constraints and present an optimization approach to minimize carbon emissions at the edge. Lastly, we present our incremental placement algorithm to our edge placement optimization in a real-world edge system. 



\subsection{\proposedsystem Overview}

\proposedsystem is a carbon-aware framework designed to reduce carbon footprint at the edge by spatially distributing workloads across edge data centers. It manages edge data centers dispersed at mesoscales, which have shown prevalently significant variations in carbon intensity (\autoref{sec:carbon_analysis}), and assumes that edge workloads
can shift across edge data centers at this scale. Edge servers are not only diverse in geolocations but also diverse in architecture and capacity, exhibiting significant differences in energy efficiency. As carbon emission is a function of energy consumption and carbon intensity of the grid, we further combine intensity variations across edge data center locations and energy efficiency differences across diverse edge servers to save carbon emissions at the edge. As a result,  \proposedsystem reduces edge carbon emissions by placing edge applications on energy-efficient edge servers with a sustainable energy supply, respecting the low-latency and resource demands. Moreover, \proposedsystem manages the power states of edge servers to reduce emissions from idle servers. 

\begin{figure}[t]
    \centering
    \includegraphics[width=1.0\linewidth]{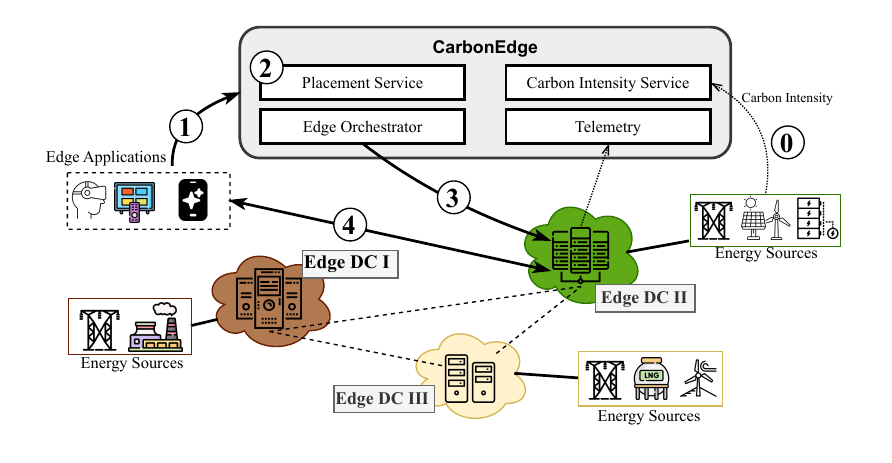}
    \caption{\proposedsystem design and exemplar workflow of placing offloading applications from IoT devices.}
    \label{fig:system_design}
    \vspace{-3mm}
\end{figure}

\autoref{fig:system_design} shows an overview of our system. The telemetry and carbon intensity services continuously collect system metrics, energy consumption, and the electricity carbon intensity of edge data centers. In addition, the carbon intensity service periodically predicts the carbon intensity of all data centers (step \Circled[inner color=black]{\textbf{0}}). When edge workloads arrive, which can be applications offloaded from resource-limited IoT or mobile devices or applications to be redeployed when an edge server fails (step \Circled[inner color=black]{\textbf{1}}), the placement service instantly decides where to allocate the workloads using a carbon-aware placement policy (step \Circled[inner color=black]{\textbf{2}}). Once the placement decisions are made, the edge orchestrator deploys the applications accordingly (step \Circled[inner color=black]{\textbf{3}}) and establishes the connections to end users (step \Circled[inner color=black]{\textbf{4}}).

\subsection{Carbon-aware Edge Placement}
\label{sec:design_problem}

The section presents the carbon-aware placement with latency constraints problem.  and our proposed policy used in \proposedsystem. Our carbon-aware policy minimizes the carbon footprint of edge data centers using an incremental optimization-driven approach. Our approach holistically integrates three factors: 1) carbon intensity variations of mesoscale edge data centers, 2) workload and diversity in energy efficiency across heterogeneous edge servers, and 3) base power usage and power proportionality of servers.


\noindent \textbf{Incremental carbon-aware placement.} Our carbon-aware placement problem aims to minimize carbon emissions from incrementally placing arriving edge applications across edge data centers while meeting strict latency requirements. Given the current power state of the servers $y_j^{curr}: j\in \mathcal{S}$ and available resource capacity $C_j^k$, our objective is to place a batch of applications $\mathcal{A}$ on $\mathcal{S}$.
First, we define two decision variables that represent the placement and power management decisions, and then define the carbon emissions associated with the placements based on these two variables. Finally, we present the carbon-aware placement problem formulation. ~\autoref{tab:notations} summarizes the notations. 

\textit{Decision variables:} Let $x_{ij} \in \{0, 1\}$ the placement of application $i \in \mathcal{A}$ on server $j \in \mathcal{S}$, and $y_j \in \{0, 1\}$ indicate whether server $j$ is powered on. These variables are subject to the following four constraint classes.

\begin{table}[tb]
\caption{Notations used in \proposedsystem.}
\label{tab:notations}
\resizebox{.95\linewidth}{!}{%
\begin{tabular}{cl}
\toprule
\multicolumn{2}{l}{\textbf{Decision variables:}} \\ 
$x_{ij} $ & True if application $i$ is placed on server $j$; False otherwise. \\ 
$y_j $ & True if server $j$ is {\em powered-on}; False otherwise. \\ 

\multicolumn{2}{l}{\textbf{Inputs:}}\\ 
$C_j^{k}$ &  Available capacity in server $j$ of type $k$ \\ 
$\bar{I_j}$ &  Average carbon intensity of server $j$ \\ 
$B_j$ & Base power usage of server $j$ when it is {\em powered-on} \\ 
$y_j^{curr}$ & Power state (on or off) of server $j$ before placement optimization \\ 
$R_{ij}^k$ &  Resource demand of type $k$ of application $i$ when running on server $j$ \\ 
$E_{ij}$ &  Energy usage of application $i$ when running on server $j$ \\ 
$L_{ij}$ & Latency between application $i$ and server $j$ \\ 
$\ell_i$ &  Latency requirement of application $i$ \\ 

\multicolumn{2}{l}{\textbf{Optimization goal:}}  \\ 
$f$ & Total carbon footprint of edge placement \\ \bottomrule 

\end{tabular}%
}
\vspace{-3mm}
\end{table}

\noindent 1) \textit{Multi-dimensional resource constraint}:  Edge servers are typically computing, storage, and networking resource-limited and are diverse in capacity and resource types. We define $C_j^k$ represents the available capacity of type $k$ on server $j$. When application $i$ runs the server $j$, its resource demands are $R_{ij}^k$. To ensure the application performance, the aggregated resource demands of applications allocated to a server must not exceed its available resources. 
{\small 
\begin{align}
    \label{eq:comp_cap}
    & \sum_i x_{ij} \cdot R_{ij}^k \leq y_j \cdot C_j^{k}, \quad \quad \quad \quad  \forall j, k 
\end{align}
}

\noindent 2) \textit{Latency constraint}: Each application $i$ comes from a certain geolocation and has a specific latency limit $L_i$. The latency from the application's source to the hosting server $L_{ij}$ must remain below this limit. 
{\small 
\begin{align}
    \label{eq:latency}
    & x_{ij} \cdot L_{ij} \leq \ell_i, \quad \quad \quad \quad \quad \quad \quad \quad \forall i, j 
\end{align}
}

\noindent 3) \textit{Placement constraint}: Each application $i$ is assigned to exactly one server. 
{\small 
\begin{align}
    \label{eq:placement}
    & \sum_j x_{ij} = 1, \quad \quad \quad \quad \quad  \quad \quad \quad \quad \quad \forall i 
\end{align}
}

\noindent 4) \textit{Power state consistency}: An active server cannot be powered off during placement (e.g., to avoid service disruption):
{\small 
\begin{align}
    \label{eq:server_keep_on}
    & y_j^{curr} \leq y_j, \quad \quad \quad \quad  \quad \quad \quad \quad \quad \forall j 
\end{align}
}
where $y_j^{curr}$ is the current power state. Additionally, assignments require active servers:  
{\small 
\begin{align}
    \label{eq:server_on}
    & x_{ij} \leq y_j, \quad \quad \quad \quad \quad  \quad \quad \quad \quad \quad \forall i, j 
\end{align}
}
\textit{Carbon emissions:} Carbon emissions from edge placement are from two main sources: application operation and server activation. Application operational emissions depend on the energy consumption of applications and the carbon intensity of hosting servers, given by $\sum_i \sum_j x_{ij} \cdot E_{ij} \cdot \bar{I_j}$, where $E_{ij}$ is the energy consumption of application $i$ on server $j$, and $\bar{I_j}$ is the average carbon intensity of the hosting server $j$. Note that carbon intensity varies over time depending on the mix of energy sources available in that area. $\bar{I_j}$ represents the average of the forecast carbon intensity values of server $j$. Server activation emissions depend on the base power $B_j$ and the carbon intensity $\bar{I_j}$, represented as $\sum_j (y_j - y_j^{curr}) \cdot B_{j} \cdot \bar{I_j}$, where $(y_j - y_j^{curr})$ represents the newly activated server. The total carbon emissions for placing applications are: 
{\small 
\begin{align}
    \label{eq:emission}
    & f =  \underbrace{\sum_i \sum_j x_{ij} \cdot E_{ij} \cdot \bar{I_j}}_{\text{Application operation}} + \underbrace{\sum_j (y_j - y_j^{curr}) \cdot B_j \cdot \bar{I_j}}_{\text{Server activation}}
\end{align}
}

\noindent \textit{Problem formulation:} The carbon-aware placement policy identifies the optimal placement $x_{ij}^*$ and power management $y_j^*$ solutions to minimize carbon emissions at the edge while meeting the latency and resource constraints. The problem is formulated as follows: 

{\small 
\begin{align}
    \label{eq:objective}
    & \min_{\substack{x_{ij}^*, y_j^*}}  \quad f  \quad \quad \quad \quad \text{s.t. Constraints 1-5 hold. }
\end{align}
}


\noindent{\bf Extensions.} The above formulation focuses on reducing carbon emissions as a primary goal while considering edge latency as a constraint. Alternatively, a multi-objective optimization can be employed to minimize both carbon emissions and latency. Additionally, other objectives like energy usage can also be optimized alongside carbon in a similar fashion. In ~\autoref{sec:eval_carbon_energy}, we  demonstrate the benefits of such a multi-objective optimization strategy, which navigates the trade-offs between carbon emissions and energy usage in edge computing.


\begin{algorithm}[t]
\scriptsize
\caption{\proposedsystem Incremental Placement Algorithm}
\label{alg:algorithm}
\raggedright
\textbf{Input:} Arriving applications $\mathcal{A}$,  Servers $\mathcal{S}$, Latency requirements $\boldmath{\ell}$, Resource demands $\mathbf{R}$, and Energy consumption $\mathbf{E}$ \\
\textbf{Output:} Placement ($\mathbf{x}$) and Power ($\mathbf{y}$) decisions.
\begin{algorithmic}[1]
\State Initialize latency matrix $\mathbf{L} \gets \emptyset$
\For{each application $a_i \in \mathcal{A}$}
\For {each server $s_j \in \mathcal{S}$}
\State $\mathbf{L}_{ij} \gets \textsc{CalculateLatency}(a_i, s_j)$ \label{ln:latency_calc} \Comment{Get application-server latency\quad }
\EndFor
\EndFor

\State $\mathcal{S}', \mathbf{L}' \gets \textsc{FilterFeasibleServers}(\mathcal{A}, \mathcal{S}, \mathbf{L}, \boldmath{\ell})$ \label{ln:preprocess} \Comment{Select servers satisfying latency limits}

\State $\mathbf{C}, B, \bar{I}, \mathbf{y}^{curr} \gets \textsc{GetServerStates}(\mathcal{S}')$ 
\Comment{Available capacities $\mathbf{C}$, base power $B$, mean carbon intensity $\bar{I}$, current power states $\mathbf{y}^{curr}$ }

\State $\mathbf{x}, \mathbf{y} \gets \textsc{SolveOptimization}(\mathcal{A}, \mathcal{S}', \boldmath{\ell}, \mathbf{R}, \mathbf{E}, \mathbf{C}, B, \bar{I} , \mathbf{y}^{curr}, \mathbf{L}')$ \label{ln:solve}
\Comment{Solve \autoref{eq:objective}}

\State $\mathbf{C}' \gets \textsc{UpdateServerStates}(\mathbf{x}, \mathbf{y}, \mathcal{C})$ \label{ln:update}
\Comment{Update servers capacity}  \\

\Return $\mathbf{x}, \mathbf{y}$
\end{algorithmic}
\end{algorithm}

\subsection{\proposedsystem Placement Algorithm}\label{sec:design_algorithm}
Our placement algorithm assumes that latency-sensitive edge applications may arrive unpredictably and need to be placed onto edge data centers in a carbon-aware manner. To achieve this, \proposedsystem employs an incremental placement strategy, executing the algorithm periodically to process newly arriving applications as a batch, to ensure carbon-efficient placement without global recomputation, as shown in \autoref{alg:algorithm}.

The algorithm places a set of arriving applications $\mathcal{A}$ (with latency requirements $\boldmath\ell$, resource demands $\mathbf{R}$, and energy profiles $\mathbf{E}$) across edge servers $\mathcal{S}$ in mesoscale data centers. First, it computes application-server latency $\mathbf{L}$ and prunes (i.e., filters out) servers exceeding latency constraints, retaining only feasible candidates $\mathcal{S}' \in \mathcal{S}$ (line 1-8). Next, it retrieves server telemetry (available capacity $\mathbf{C}$, base power $B$, current power state $y^{curr}$) and  mean carbon intensity $\bar{I}$ (line 9). Then, it solves the optimization in ~\autoref{eq:objective} using the latency-compliant server set $\mathcal{S}'$ to ensure traceability. 
Placing incoming applications in small batches in real-time can be done efficiently---our  result in ~\autoref{sec:eval_overhead} shows that  incremental placement of a batch of 50 newly arriving applications across 400 servers completes in 3 seconds (line 10). Finally, \proposedsystem commits the resource allocation and power state transition, updating server states for the next iteration (line 11).

\section{\proposedsystem Implementation}
\label{sec:implementation}
This section describes the implementation of \proposedsystem (See ~\autoref{fig:system_design}) on top of Sinfonia~\cite{satyanarayanan2022sinfonia}, a Kubernetes based open-source orchestrator for edge-native applications. Our implementation adds $\sim$4k SLOC to the  Sinfonia system and is available as open source at https://github.com/umassos/CarbonEdge. 


\subsection{\proposedsystem Prototype}
\noindent Our \proposedsystem consists of the following components that we added to Sinfonia.

\noindent\textbf{Telemetry Service}:
Our telemetry service is integrated into Sinfonia’s telemetry, where it collects static (e.g., location and IP address) attributes and real-time  (e.g., utilization) metrics. Real-time metrics are collected based on the Prometheus monitoring stack\cite{Prometheus}. We augment Sinfonia's monitoring with the following metrics:
\begin{enumerate}[leftmargin=*]
    \item \textbf{Power Monitoring}: We measure the power consumption of CPU servers using RAPL~\cite{david2010rapl}, and we leveraged Prometheus's DGCM exporter for GPUs~\cite{nvidia_dcgm_exporter_github}. 

    \item \textbf{Carbon Intensity}: We integrate a carbon intensity service that replays historical traces from Electricity Maps~\cite{electricity-map} and uses the traces to provide real-time and forecast carbon intensity.

    \item \textbf{Carbon Monitoring}: We implement carbon monitoring based on energy usage and the carbon intensity of the selected edge sites, where we account for the base power (if the server is turned on) and applications' energy usage. 
    
    \item \textbf{End-to-end latency}: In addition to latency across sites, we recorded end-to-end latency between users and their deployed applications.

\end{enumerate}

\noindent\textbf{Profiling Service}: 
We implement an application profiling service that collects the application's performance metrics, such as latency, power consumption, resource demands, and other crucial information, to make accurate placement decisions across available resources. Our profiling service can be replaced with performance models that statically analyze the applications and predict the latency and energy consumption~\cite{paleo, cai2017neuralpower}.

\noindent\textbf{Placement Service}:
Lastly, we implement our placement policy (\autoref{alg:algorithm}) on Sinfonia as a matching policy. The placement policy utilizes the system's real-time metrics, static attributes of different edge sites, and workload profiles to determine optimal placements and server activation.
Our implementation batches deployment requests (e.g., every 5 minutes) and solves the optimization problem per application batch using the Google OR-Tools~\cite{ortools}. We demonstrate the effectiveness of our approach in ~\autoref{sec:eval_overhead}. 
After computing the placement decisions, we utilize Sinfonia's orchestration capabilities to initiate the deployment sequence (Sinfonia RECIPE), which contains the necessary Kubernetes deployment files and helm charts, to the destination servers or activate servers if necessary, and inform the client(s) of the destination's address. Note that although Sinfonia and our system are packed with fault-tolerance and reconfiguration capabilities, evaluating them is beyond the scope of this paper.

\subsection{\proposedsystem Edge Simulator}
In addition to the prototype of \proposedsystem, we developed a simulator for larger-scale evaluations that is not feasible using an edge testbed.
Our simulator supports simulating diverse edge settings with dynamic workloads and heterogeneous servers. This simulator represents the components of Sinfonia and follows the same decision process and metrics, where we implement our proposed carbon-aware placement policy and other baseline policies using Google OR-Tools~\cite{ortools}. The \proposedsystem simulator is implemented in Python using $\sim$2k SLOC.

\section{Evaluation}
\label{sec:evaluation}

In this section, we evaluate the performance of \proposedsystem using real experiments and large-scale simulations. We start with evaluating \proposedsystem in mesoscale edge deployments, showcasing its efficiency in reducing carbon emissions on a regional scale. Next, we extend our analysis to continental-scale edge data centers (e.g., a Content Delivery Network, CDN), highlighting that the benefits of saving carbon emissions with granular carbon intensity are commonplace. In doing so, we address the following questions:

\begin{enumerate}[leftmargin=*]
    \item {\em What are the potential carbon savings of spatial shifting for an edge provider with multiple regional edge data center locations?}
    \item {\em How can a CDN exploit mesoscale variations for carbon-aware edge hosting across a large network of edge locations? How do latency limits affect potential carbon savings?}
    \item {\em How do seasonal variations, demand, and capacity affect carbon savings and placement decisions?}
    \item {\em How does the heterogeneity in edge resources impact savings? What are the carbon-energy trade-offs in these settings?}
\end{enumerate}

Next, we detail our real-world datasets, experimental settings, baselines, and evaluation metrics.

\begin{figure}[t]
    \centering
    \includegraphics[width=0.6\linewidth]{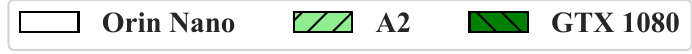}\\
    \subfloat[\centering Energy]{\includegraphics[width=0.3\linewidth]{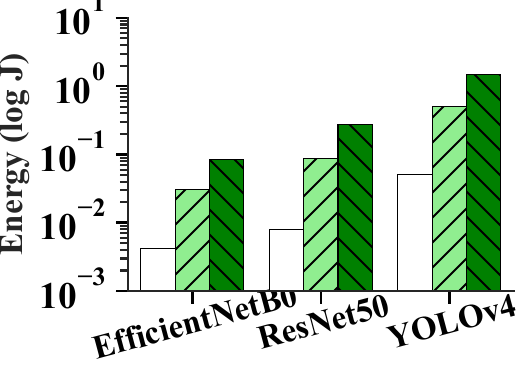}%
    \label{fig:models_profile_energy}%
    }%
    \subfloat[\centering GPU Memory]{
    \includegraphics[width=0.3\linewidth]{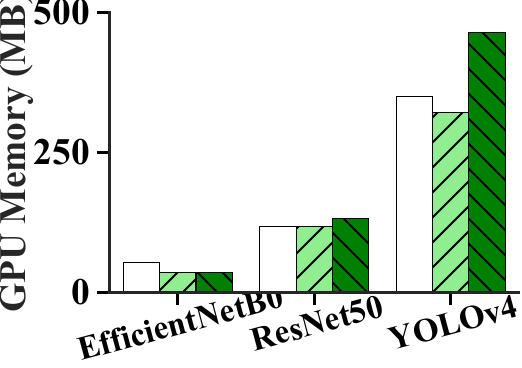}
    \label{fig:models_profile_memory}
    }%
    \subfloat[\centering Inference Time]{
    \includegraphics[width=0.3\linewidth]{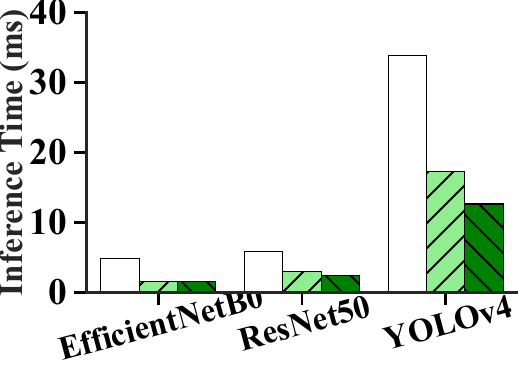}
    \label{fig:models_profile_latency}
    }%
    \caption{Energy consumption, memory usage, and inference time of ML workloads across devices.}
    \label{fig:models_profile}
    \vspace{-5mm}
\end{figure}

\subsection{Experimental Methodology}
\subsubsection{Real World Traces.}\label{sec:real_world_traces}\hfill\\
This section describes our real-world traces and how we combined them in our evaluations.

\noindent{\bf Carbon Intensity Traces.}
We utilize the carbon intensity traces for 2023 from Electricity Maps ~\cite{electricity-map}. The trace contains the hourly carbon intensity, measured in \carbonunit, for 148 carbon zones worldwide, including 54 and 45 in the US and Europe, respectively. Electricity Maps define each zone according to the structure of the regional electricity grid. For example, the results show that the area of carbon zones can be as small as 123.73 $km^2$ (Tallahassee, Florida).

\noindent{\bf Latency Traces.} To incorporate realistic network latencies between edge data centers. 
We used the round-trip latency traces from WonderNetwork~\cite{wonder-proxy-2020}, which provides ping times (in milliseconds) between 246 major cities worldwide. The data covers 64 cities in the US and 64 cities in Europe. Each city is associated with longitude and latitude coordinates. The data highlights that in the US, latency can range from 0.93 ms to 184.67 ms, with an average of 43.17 ms, while in Europe, it ranges from 1.12 ms to 156.74 ms, with an average of 36.94 ms. 

\noindent{\bf Edge Workloads.}
We utilize two types of compute-intensive edge workloads: a CPU-based edge application that emulates edge sensor data processing and a GPU-based model-serving application that emulates edge AI inference. The CPU-based application is implemented using Python and numpy v1.26, while the model-serving application uses TensorRTv10.2 and CUDA 12.1.
~\autoref{fig:models_profile} depicts the three selected models that cover different tasks and resource requirements: EfficientNetB0~\cite{efficient-net}, ResNet50~\cite{resnet}, and YOLOv4~\cite{bochkovskiy2020yolov4}. ~\autoref{fig:models_profile_energy} highlights the diversity of our workloads, where energy consumption can reach 45$\times$  across models in the same device, and 2$\times$ across devices for the same model. Similarly, ~\autoref{fig:models_profile_memory} shows that memory also differs across models and devices. We evaluate the effect of heterogeneity in ~\autoref{sec:eval_hetero}. Unless mentioned otherwise, we assume a round-trip network latency constraint of 20 ms ($\sim$500km) in our experiments. 


\noindent{\bf Edge Data Centers.}
To emulate real-world edge deployment, we utilize Akamai CDN traces, which include the location of edge data centers globally identifiable by their coordinates. 


\noindent{\bf Integrating Traces.}
Since the availability of granular data differs between traces, we integrate the traces using the following steps:
\begin{enumerate}[leftmargin=*]
    \item We map each data center in the Akamai trace to its corresponding carbon intensity zone using its coordinates.
    \item We compute the cross-data center latency by mapping each data center to the nearest city. We assume that users exhibit the same latency as their original edge data center.
    \item We limit our evaluations to Akamai CDN edge data centers where the carbon intensity and latency traces are available. 
    \item In the case of multiple data centers in the same city, we combine them into a single data center.
\end{enumerate}

\subsubsection{Experimental Setup} \hfill\\
We evaluate our \proposedsystem under two deployment scenarios:

\noindent {\bf Mesoscale Regional Edge Deployment.} We first evaluate the performance of \proposedsystem in mesoscale edge deployments. In our experiments, we use eleven servers to emulate a mesoscale edge network, which comprises five edge data centers distributed across five cities. Each data center is represented by a server and associated with an end device, also represented by a server, for issuing application placement and service requests. \proposedsystem operates on a separate server to prevent any interference. The eleven servers are Dell PowerEdge R630, each equipped with a 40-core Xeon E5-2660v3 CPU, 256GB of memory, and a 1Gb/s network connection. Additionally, each edge server contains an NVIDIA A2 GPU that has 1280 CUDA cores, 16GB of memory, and 60W of maximum power consumption, enabling us to evaluate \proposedsystem on a GPU cluster. Lastly, we used a workload generator based on Locust\footnote{Locust: \url{https://locust.io/}} and used the Linux traffic control tool ($tc$\footnote{ \url{https://man7.org/linux/man-pages/man8/tc.8.html}}) to emulate network latency across edge data centers.

\noindent {\bf Continental-scale CDN Edge Deployment.}
In addition to mesoscale evaluations, we show how mesoscale variations can help decarbonize CDN edge deployment that spans an entire continent (e.g., Akamai CDN and AWS Local Zones). In this case, we utilize trace-driven simulations to evaluate the year-long global behavior of the CDN across the US and Europe.
In ~\autoref{sec:eval_hetero}, we show the impact of heterogeneity across data centers, where we profiled the ML workloads mentioned above on NVIDIA A2 (1280 CUDA cores, 16GB memory, 60W), NVIDIA Jetson Nano (1024 CUDA cores, 8GB memory, and 15W ), and NVIDIA GTX 1080 (2560 CUDA cores, 8GB memory, 180W).   


\subsubsection{Baselines} \hfill\\
We evaluate \proposedsystem against multiple baselines.

\begin{enumerate}[leftmargin=*]
\item {\bf \latencyaware}: This policy allocates workloads to the nearest edge data centers to minimize latency overhead, a strategy commonly employed in edge computing~\cite{yi2017:Lavea}.

\item {\bf \energyaware}: This policy distributes workloads to energy-efficient edge data centers to decrease energy consumption~\cite{Goudarzi21Placement, Li2018_energy_placement}. We implemented this policy by minimizing energy usage while adhering to latency and resource constraints. 
 
\item {\bf \intensityaware}: This policy greedily assigns workloads to the greenest edge data centers with the lowest carbon intensity values while respecting the latency and resource constraints. 


\end{enumerate}

\subsubsection{Evaluation Metrics} \hfill\\
We evaluate \proposedsystem with three key metrics: Carbon Emissions, Response Time, and Energy Consumption, where we report absolute values as well as carbon savings (\%), {\em round-trip} latency increases (ms), and energy consumption compared to the \latencyaware baseline.

    


\begin{figure}[t]
    \centering
    \includegraphics[width=0.9\linewidth]{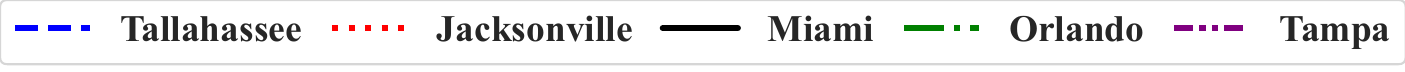}\\
    
    \subfloat[\centering Carbon Intensity]{{
        \includegraphics[width=0.3\linewidth]{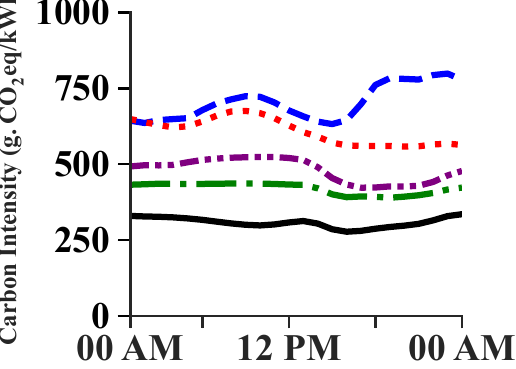}}
        \label{fig:carbon_emission_ts_intensity}
    }%
    \subfloat[\centering Latency-aware]{{
        \includegraphics[width=0.3\linewidth]{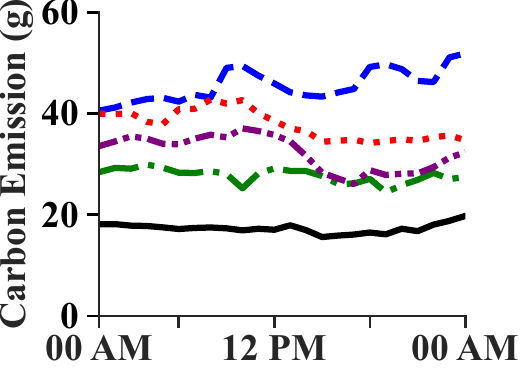}}
        \label{fig:carbon_emission_ts_agnostic}
    }%
    \subfloat[\centering CarbonEdge]{{
        \includegraphics[width=0.3\linewidth]{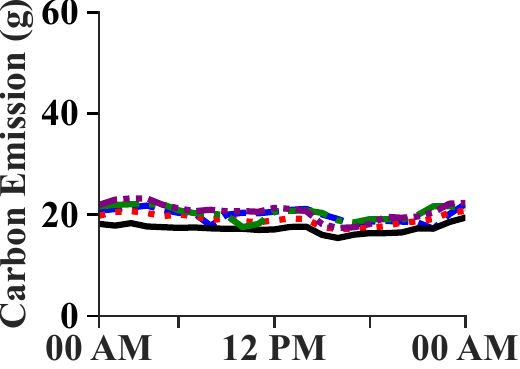}}
        \label{fig:carbon_emission_ts_aware}
    }%
    \caption{Carbon intensity and emissions across edge data centers in Florida.}
    \label{fig:carbon_emission_ts}
\end{figure}

\begin{figure}[t]
    \centering
    \includegraphics[width=1\linewidth]{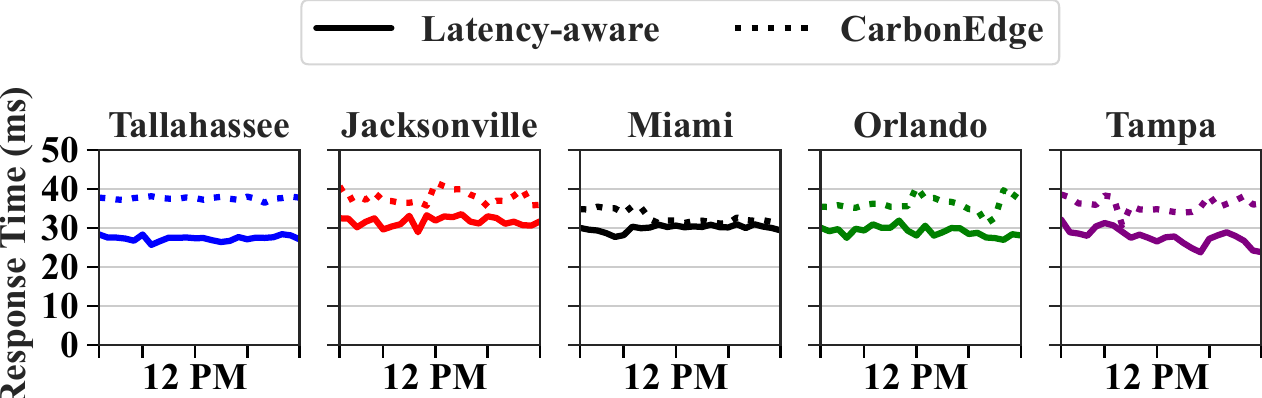}
    \caption{End-to-end response times of applications across edge data centers in Florida.}
    \label{fig:response_time_ts}
    \vspace{-4mm}
\end{figure}


\subsection{Mesoscale Evaluation}\label{sec:eval_mesoscale}
We start by evaluating the performance of the \proposedsystem prototype for two regional deployments (in Florida and Central Europe) over a 24-hour period. We compare the performance of \proposedsystem to the \latencyaware baseline. \autoref{fig:carbon_emission_ts} illustrates the carbon intensity and emissions of the CPU-based application across five zones within the Florida region. 
As explained in ~\autoref{sec:carbon_analysis}, zones in Florida exhibit high variations (see ~\autoref{fig:carbon_emission_ts_intensity}). ~\autoref{fig:carbon_emission_ts_agnostic} shows the carbon emissions of the \latencyaware policy, which highly resemble each zone's carbon intensity in \autoref{fig:carbon_emission_ts_intensity}. In contrast, \autoref{fig:carbon_emission_ts_aware} shows how \proposedsystem places all applications in the greenest zone (Miami), resulting in an equivalent 20-23 \emissionunit of emissions for all applications.
\autoref{fig:response_time_ts} lists how the response time changes between \latencyaware and \proposedsystem across different data centers. As expected, increases in response time are limited due to the proximity of different data centers, where the response time increases are <10.1 ms, with an average increase of 6.61 ms.  

\begin{figure}[tb]
    \subfloat[\centering Carbon Emissions ]{\includegraphics[width=0.55\linewidth]{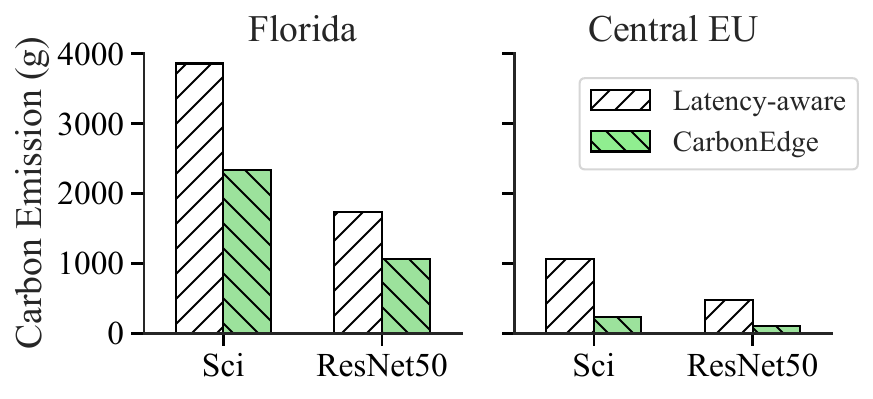}%
    \label{fig:comparion_regions_real_carbon}%
   }
    \subfloat[\centering Latency Increases]{
    \includegraphics[width=0.4\linewidth]{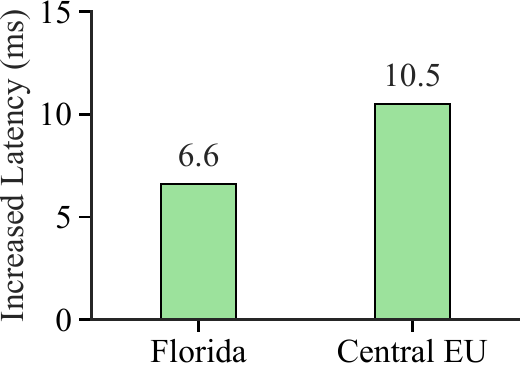}
    \label{fig:comparion_regions_real_latency}
    }%
    \caption{Performance of \proposedsystem across applications, policies, and locations.}
\label{fig:comparion_regions_real}
\end{figure}

~\autoref{fig:comparion_regions_real} depicts the aggregate emissions and latency overheads across regions for the CPU-based and GPU-based applications (ResNet50). As shown in ~\autoref{fig:comparion_regions_real_carbon}, carbon emissions vary significantly across regions and applications. For instance, in Central Europe, total carbon emissions are reduced by up to 2.6$\times$ and 10.3$\times$ for the \latencyaware and \proposedsystem policy compared to those in Florida, where total emission is a function of the average carbon intensity, as highlighted in earlier research~\cite{hanafy2023carbonscaler}.
The figure also highlights how power consumption impacts total carbon emissions. For instance, the GPU-based application emits 54.7\% less carbon, which is proportional to the differences in power consumption between the CPU-based application and the GPU-based application. However, since the proposed system implements the same placement decisions apart from the application requirements, the carbon savings and latency increases remain consistent. Overall, \proposedsystem lowers carbon emissions by 39.4\% in Florida and 78.7\% in Central EU. Meanwhile, response time increased by 6.6 ms for Florida and 10.5 ms for Central EU (shown in ~\autoref{fig:comparion_regions_real_latency}).


\noindent \textit{\textbf{Key Takeaways.} 
In mesoscale edge settings, \proposedsystem can highly optimize the carbon emissions resulting in 39.4\% and 78.7\% carbon savings for Florida and Central EU, respectively. 
}

\subsection{Mesoscale Evaluation for a CDN}\label{sec:eval_global}
In this section, we evaluate \proposedsystem for deploying edge applications in a continental scale CDN using a year-long simulation. 
We focus our evaluation on US 
and European CDN edge data centers only 
since fine-granular carbon intensity information for other continents was unavailable.  
In a CDN, edge applications arrive at edge data centers over time. Each edge application comes from a specific zone, and its placement is limited to a subset of edge data centers within a certain radius (or latency) of that site.



\subsubsection{Year-long Performance Evaluation}\label{sec:eval_understanding}
\begin{figure}[tb]
    \subfloat[\centering Carbon  ]{
    \includegraphics[width=0.26\linewidth]{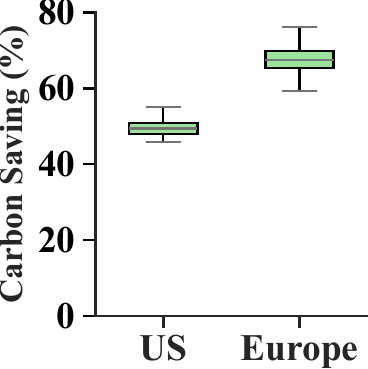}
    \label{fig:large_scale_carbon}
    }%
    \subfloat[\centering Latency  ]{
    \includegraphics[width=0.26\linewidth]{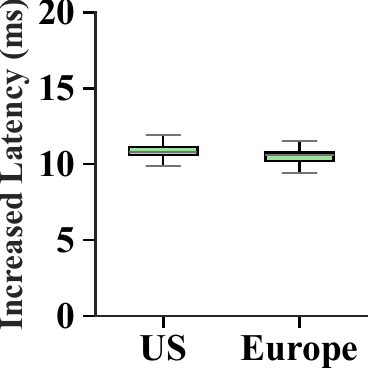}
    \label{fig:large_scale_latency}
    }%
    \subfloat[\centering Load Distribution]{
    \includegraphics[width=0.42\linewidth]{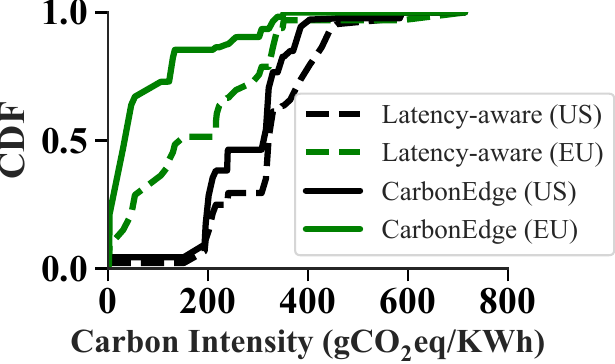}
    \label{fig:large_scale_cdf}
    }%

    \caption{Carbon savings, latency increases, load distribution in the US and Europe.}
    \label{fig:large_scale}
\end{figure}

\hfill\\
\autoref{fig:large_scale} presents the year-long performance of \proposedsystem in terms of carbon savings and latency overheads when considering a latency limit of 20 ms ($\sim$500 km). As shown, \proposedsystem reduces carbon emissions by 49.5\% in the US and 67.8\% in Europe. Meanwhile, the average latency increases by 10.8 ms in the US and 10.5 ms in Europe. We note that Europe experiences higher carbon savings as the European data centers reside in greener zones, and the carbon intensity variations across these data centers are larger than those in the US. In addition, ~\autoref{fig:large_scale_cdf} illustrates the workload shifting with \latencyaware and \proposedsystem in the US and Europe. The results highlight that \proposedsystem shifts workloads toward low-carbon edge locations. For instance, compared to the \latencyaware baselines, \proposedsystem increases application execution at 200\emissionunit by 40\% and 33.9\% for the US and Europe, respectively. Moreover, the figure highlights examples where edge data centers do not have any of their load shifted as they are far away from other greener regions. For instance, in the US, the edge data center in Salt Lake City, Utah, does not offload any of its load.



\noindent \textit{\textbf{Key Takeaways.} 
By shifting the demand towards low carbon zones, \proposedsystem decreases carbon emissions by 49.5\% and 67.8\% for the US and Europe, respectively, while increasing the round-trip latency by less than 11 ms.
}

\subsubsection{Impact of Latency Tolerance}\label{sec:eval_latency}
\begin{figure}[tb]
    \centering
    \subfloat[\centering Carbon Savings ]{\includegraphics[width=0.4\linewidth]{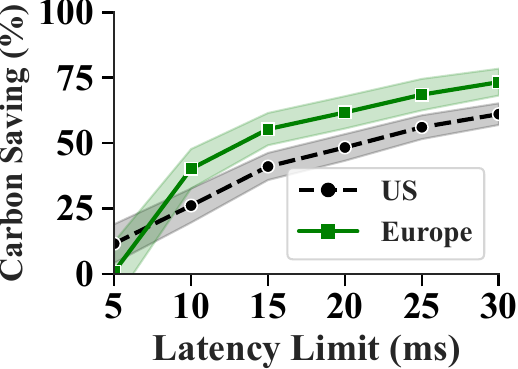}%
    \label{fig:latency_effect_carbon}%
    }%
    \quad
    \subfloat[\centering Latency Increases ]{
    \includegraphics[width=0.4\linewidth]{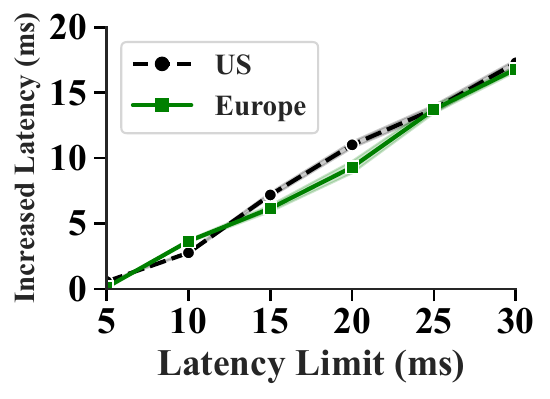}
    \label{fig:latency_effect_overhead}
    }%
    \caption{Effect of latency tolerance on carbon savings and latency increases.} 
    \label{fig:latency_effect}
    \vspace{-3mm}
\end{figure}

\hfill\\
Carbon savings are a function of placement flexibility, where applications with no latency requirements can be placed in locations with zero carbon intensity~\cite{sukprasert2024limitations}. However, in practice, edge applications have tight latency requirements. ~\autoref{fig:latency_effect} depicts the carbon savings and latency overheads across different round-trip latency limits in the US and Europe. As shown in ~\autoref{fig:latency_effect_carbon}, allowing a round-trip latency tolerance of 10 ms can yield 28\% and 44.8\% carbon savings in the US and Europe, respectively, while raising the latency limit to 20 ms can increase these carbon savings by 23\% and 23.4\%. These increasing carbon savings come from placing more workload in greener regions that meet the latency limits. 
Moreover, the figure emphasizes that increasing latency limits leads to diminishing returns. For example, in Europe, increasing the latency limit from 5 ms to 10 ms increases savings by 43.8\%, whereas increasing the limit from 25 ms to 30 ms only yields an extra 4\% savings. ~\autoref{fig:latency_effect_overhead} shows that performance overheads increase linearly with rising latency limits. Importantly, the results indicate that the benefits consistently outweigh the overheads. For instance, the figure demonstrates that the \proposedsystem reduces carbon emissions by 74.7\% while incurring only a 17.2 ms increase in round-trip latency.

\noindent \textit{\textbf{Key Takeaways.} 
For a 10 ms increase in latency, \proposedsystem derives 28\% and 44.8\% carbon savings in the US and Europe, respectively. Notably, the results demonstrate that the benefits constantly outweigh the overheads, where \proposedsystem reduces carbon emissions by 74.7\% while incurring only a 17.2 ms increase in round-trip latency.
}

\begin{figure*}[t]
    \raggedleft
    \quad\includegraphics[width=0.44\linewidth]{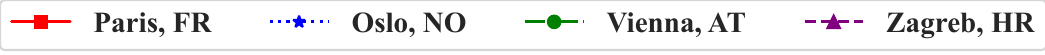}\quad \quad \quad \\
    \centering
    \subfloat[\centering Carbon Savings]{\includegraphics[width=0.22\linewidth]{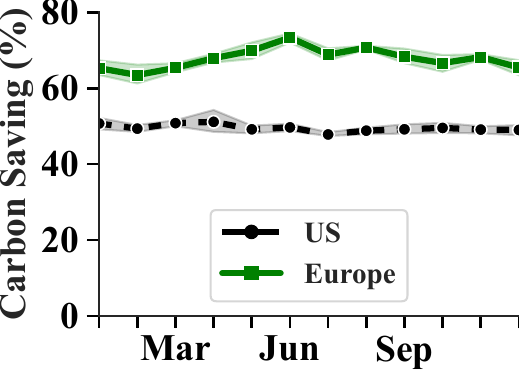}%
    \label{fig:global_monthly_carbon}%
    }%
    \quad
    \subfloat[\centering Latency Increases]{
    \includegraphics[width=0.22\linewidth]{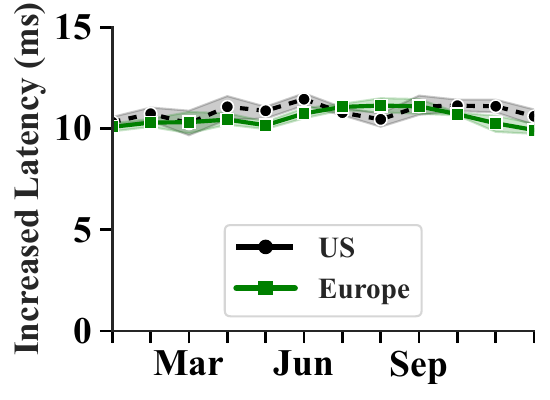}
    \label{fig:global_monthly_latency}
    }%
    \quad
    \subfloat[\centering Carbon Intensity]{
    \includegraphics[width=0.22\linewidth]{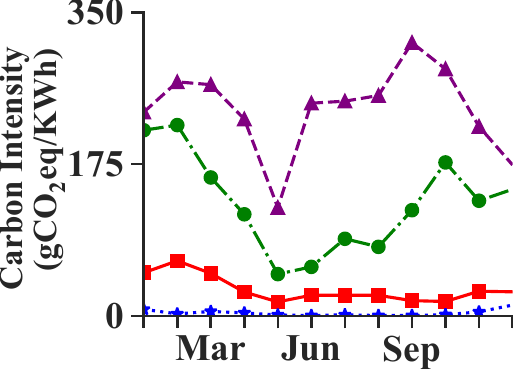}
    \label{fig:monthly_changes_intensity}
    }%
    \quad
    \subfloat[\centering Placements]{
    \includegraphics[width=0.22\linewidth]{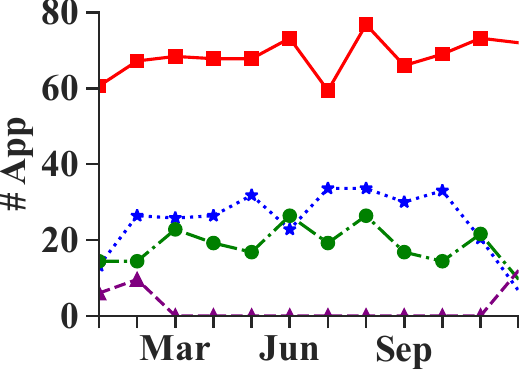}
    \label{fig:monthly_changes_placement}
    }%
    \caption{Effect of seasonality on carbon savings and latency overhead across the US and Europe.} 
    \label{fig:global_monthly}
\end{figure*}


\subsubsection{Impact of Seasonality}\label{sec:eval_seasonality}\hfill\\
To better understand the effect of seasonality on spatial decisions, ~\autoref{fig:global_monthly} illustrates the fluctuations in carbon savings and latency increases over 12 months in the US and Europe. ~\autoref{fig:global_monthly_carbon} shows that carbon savings in the US exhibit minimal changes, with a maximum difference of 3.3\% in carbon savings (i.e., between July and April). In contrast, carbon savings vary significantly in Europe, resulting in a 9.9\% difference between February and June. Furthermore, ~\autoref{fig:global_monthly_latency} indicates that latency overheads see only slight changes, with only 1.2 ms differences in both the US and Europe.

~\autoref{fig:monthly_changes_intensity} and \autoref{fig:monthly_changes_placement} further detail how seasonal variations in carbon intensity impact placement decisions in \proposedsystem, demonstrating the need for migrating long-running applications across regions. As shown in ~\autoref{fig:monthly_changes_intensity}, the changes in carbon intensity vary significantly between locations. For example, Zagreb, HR exhibits a 102 \carbonunit difference between April and May, whereas Oslo, NO only sees a 2.4 \carbonunit change. Reflecting these changes in carbon intensity, the number of applications assigned to these data centers fluctuates greatly. As indicated in \autoref{fig:monthly_changes_placement}, the number of applications assigned to Paris varies by 1.3$\times$ between July and August, while in Oslo, it changes by 3$\times$ between December and November. Additionally, the figure indicates that variations in demand  might be reflected in the carbon intensity of nearby areas rather than in the region itself. For example, Oslo, which exhibits the most significant fluctuation in application assignments, demonstrates little change in carbon intensity. Conversely, Vienna, with the largest shifts in carbon intensity, sees only minor changes in the volume of assigned applications.
 
 
 %

\noindent \textit{\textbf{Key Takeaways.} 
The seasons' changes in carbon intensity highly affect the carbon savings that change by up to 10\% across months. The intertwined relations between regions change across seasons, resulting in up to 3$\times$ change in resource allocation.
}


\begin{figure}[tb]
    \centering
    \includegraphics[width=0.5\linewidth]{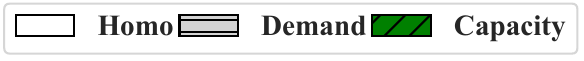} \\
    \subfloat[\centering Carbon Savings]{\includegraphics[width=0.4\linewidth]{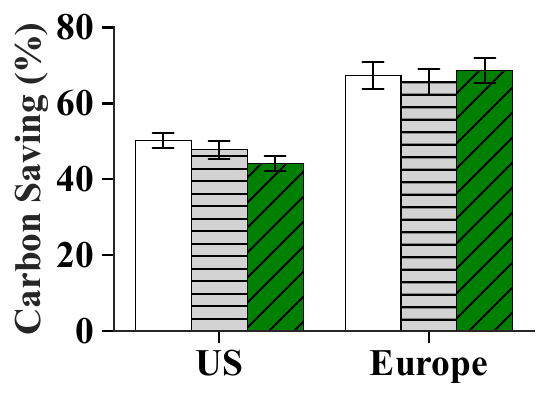}%
    \label{fig:workload_capacity}%
    }%
    \quad
    \subfloat[\centering Increased Latency]{
    \includegraphics[width=0.4\linewidth]{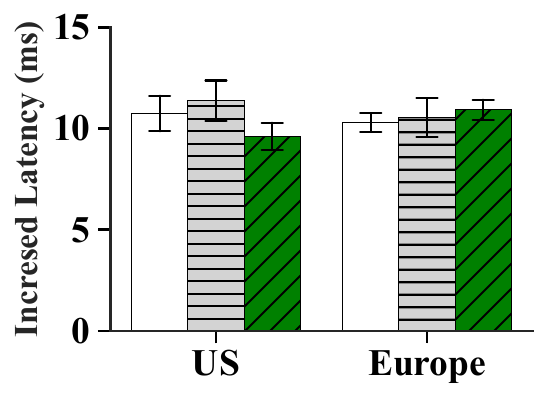}
    \label{fig:workload_increased_latency}
    }%
    \caption{Effect of demand and capacity. }
    \label{fig:demand_capacity}
    \vspace{-3mm}
\end{figure}

\subsubsection{Impact of Demand and Capacity}\label{sec:eval_demand_capacity}\hfill\\
Carbon savings from edge placements are affected by regional demand and resource capacity. This section evaluates the impact of demand and capacity using population data as a proxy for such differences. Our intuition behind this is that locations with high populations typically have high demand. Similarly, edge providers tend to increase their capacities near them.
~\autoref{fig:demand_capacity} demonstrates the impact of demand and capacity variations on carbon savings. The demand represents the case where the workload across data centers is proportional to the population across cities, while keeping the capacity fixed, while the capacity scenario changes the capacity distribution according to the population density, while keeping the demand fixed. Lastly, for reference, we add the homogeneous scenario (labeled as Homo), where the demand and capacity are constant across data centers.  
As shown, in the US, changes in demand and capacity can limit the flexibility to do spatial shifting and decrease carbon savings. For instance, changes in capacity per the population reduce carbon emissions by 6\%. This is because high carbon intensity locations (e.g., FL) have no nearby green regions to shit workloads. In contrast, in Europe, the population is more evenly distributed within the data centers we utilize, where carbon savings changes are <1.6\% and latency changes by <0.6 ms.

\noindent \textit{\textbf{Key Takeaways.} 
Changes in demand and capacity can impact carbon savings based on the carbon intensity of their origin. 
}

\subsubsection{Impact of Heterogeneity}\label{sec:eval_hetero}\hfill\\
Heterogeneity is an inherent property of edge computing that appears in applications and systems~\cite{HeteroEdge}. In this section, we evaluate the performance of \proposedsystem with diverse edge applications and heterogeneous edge servers and compare it to three baselines: \latencyaware, \energyaware, and \intensityaware.

\begin{figure}[tb]
    \centering
    \includegraphics[width=1.0\linewidth]{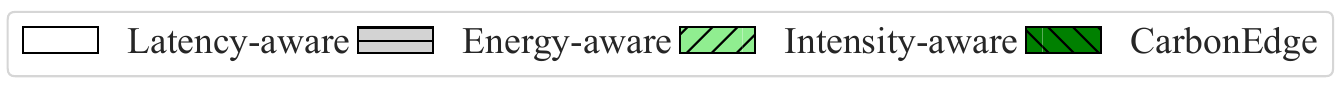} \\ 
    \subfloat[\centering Carbon Emissions (log t)]{\includegraphics[width=0.485\linewidth]{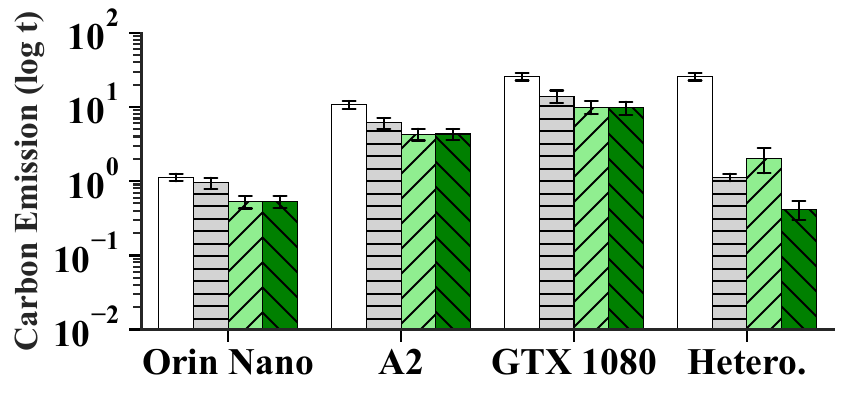}%
    \label{fig:server_heteroginity_carbon}%
    }%
    \hfill
    \subfloat[\centering Energy Consumption]{
    \includegraphics[width=0.485\linewidth]{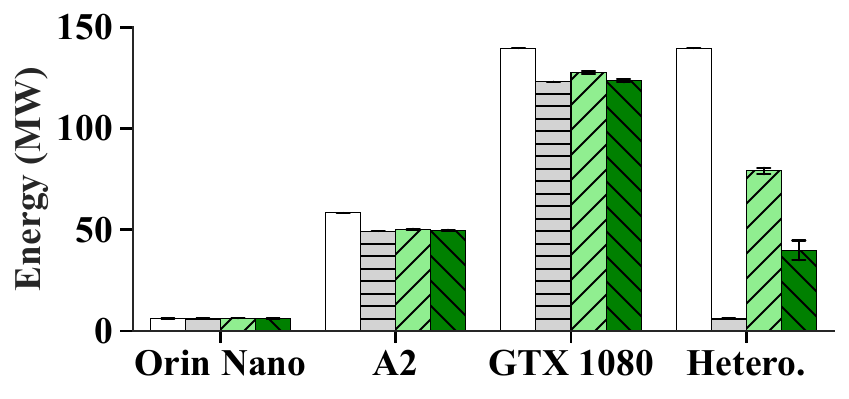}
    \label{fig:server_heteroginity_energy}
    }%
    \caption{Carbon emissions and energy consumption across workloads on heterogeneous resources. }
    \label{fig:server_heteroginity}
    \vspace{-5mm}
\end{figure}

~\autoref{fig:server_heteroginity} illustrates the carbon emissions and energy consumption of a mix of applications, including EfficientNetB0, ResNet50, and YOLOv4, across three different resources (Orion Nano, A2, and GTX 1080) and a mix of them (labeled as Hetero.). ~\autoref{fig:server_heteroginity_carbon} shows \energyaware, \intensityaware, and \proposedsystem can reduce carbon emissions compared to \latencyaware. The figure highlights the energy efficiency of different hardware, showing that serving the same load using Orion Nano uses 95.6\% less energy than GTX 1080. However, \proposedsystem achieves 53\% and 62\% carbon reductions for Orion Nano and GTX 1080, respectively.  This is because, although the GTX 1080 has higher energy consumption, its low inference time (see~\autoref{fig:models_profile_latency})  can enlarge the potential of spatial shifting, allowing requests to be offloaded to low-carbon locations.  Importantly, when considering heterogeneous resources, \proposedsystem can further reduce carbon emissions by interplaying the differences in energy efficiency, carbon intensity, and processing speed, decreasing carbon emissions by 98.4\%, 79\%, and 63\% reductions compared to the \latencyaware, \intensityaware, and \energyaware baselines, respectively. ~\autoref{fig:server_heteroginity_energy} 
highlights the carbon-energy trade-off, where carbon-aware placement increases the total energy consumption.  Compared to the energy-aware placement, \intensityaware and \proposedsystem can use 12$\times$ and 5.5$\times$ more energy.

\noindent \textit{\textbf{Key Takeaways.} 
By interplaying the differences in energy efficiency, carbon intensity, and processing speed,  \proposedsystem can reduce carbon emissions by 98\%, 79\%, and 63\% compared to the \latencyaware, \intensityaware, and \energyaware baselines, respectively. 
}

\subsection{Navigating the Carbon-Energy Trade-off}\label{sec:eval_carbon_energy}
Despite the importance of carbon emissions from edge computing, it is crucial not to ignore the trade-off between lowering carbon emissions and energy consumption, as energy typically incurs a monetary cost. 
To understand the breadth of the carbon-energy trade-off, we augment the optimization objective in \autoref{eq:objective} as follows:
{\small 
\begin{align}
    \label{eq:carbon_energy_tradeoff}
    & \min_{\substack{x_{ij}^*, y_j^*}} \quad \alpha \cdot p + (1-\alpha) \cdot f
\end{align}
}

where $p$ is the total energy consumption, $f$ is the total carbon footprint and $\alpha$ is a weighting factor, where $\alpha=0$ results in the vanilla \proposedsystem policy, while $\alpha=1$ is the \energyaware policy.
Note that to navigate the trade-off seamlessly, we normalize the carbon intensity and energy consumption between $ [0, 1]$ using min-max normalization.

\begin{figure}[tb]
    \centering
    \subfloat[\centering Low Utilization ]{\includegraphics[width=0.48\linewidth]{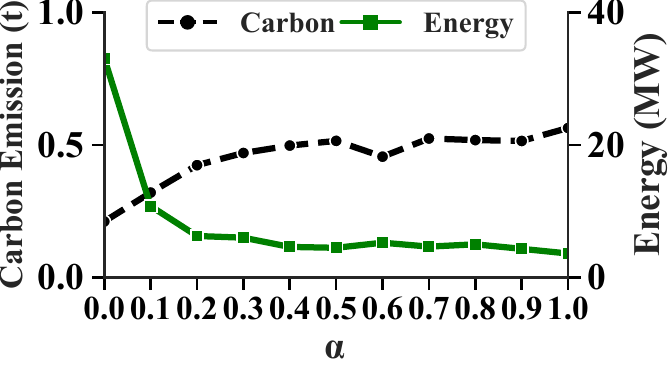}%
    \label{fig:carbon_energy_low}%
    }%
    \hfill
    \subfloat[\centering High Utilization ]{
    \includegraphics[width=0.48\linewidth]{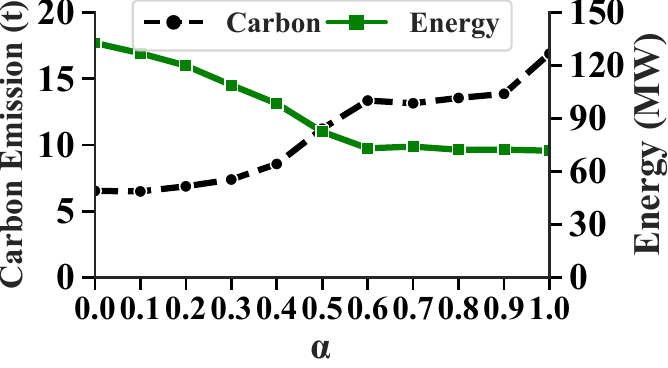}
    \label{fig:carbon_energy_high}
    }%
    \caption{\proposedsystem with carbon-energy trade-offs.} 
    \label{fig:carbon_energy}
    \vspace{-7mm}
\end{figure}

~\autoref{fig:carbon_energy} depicts how changing $\alpha$ can affect carbon emissions and energy consumption within two scenarios: low and high resource utilization. As shown, as the utilization increases, the magnitude of carbon emissions and energy highly increases, where carbon emissions and energy consumption increase by 15.5$\times$ and 20$\times$ from the low to the high utilization scenarios. 
Moreover, in both cases, \proposedsystem significantly reduces carbon emissions, where it reduces carbon emissions by 98.4\% and 90.5\% for the low and high utilization compared to the \latencyaware, respectively. Nonetheless, in both cases, carbon-efficient placement ($\alpha=0$) increases energy consumption compared to energy-efficient placement ($\alpha=1$) by 9.1$\times$ and 1.84$\times$ for the low and high utilization scenarios, respectively.
Lastly, the figure highlights that, in both cases, a balance point exists where carbon reductions come at a lower energy cost. For instance, in the low utilization scenario (\autoref{fig:carbon_energy_low}), using $\alpha=0.1$ allows \proposedsystem to retain 97.5\% of its carbon savings while decreasing energy consumption by 67\%. In contrast, in the high utilization scenario (~\autoref{fig:carbon_energy_high}), the trade-off is more prominent, where using $\alpha=0.5$ allows \proposedsystem to retain 83.7\% of its carbon savings while increasing energy by 15\%.

\noindent \textit{\textbf{Key Takeaways.} 
The inherent carbon-energy trade-off is pronounced in heterogeneous edge settings. By augmenting the objective function with energy-awareness, \proposedsystem can maintain 97.5\% of its carbon savings while decreasing energy consumption by 67\%.
}

\subsection{System Overhead}\label{sec:eval_overhead}
\begin{figure}[t]
    \subfloat[\centering Number of Servers]{
    \includegraphics[width=0.24\linewidth]{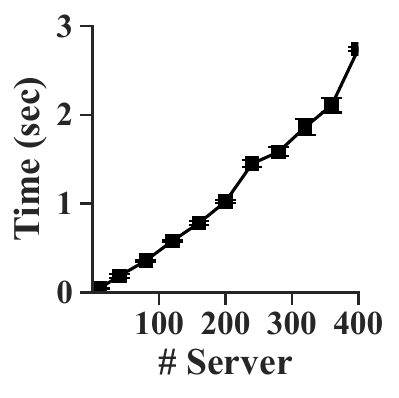}
    \label{fig:server_time}
    \includegraphics[width=0.24\linewidth]{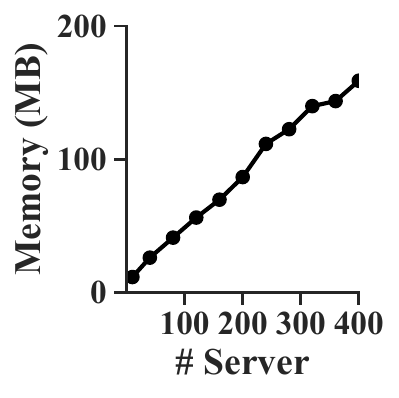}
    \label{fig:server_mem}
    }%
    \subfloat[\centering Number of Applications]{\includegraphics[width=0.24\linewidth]{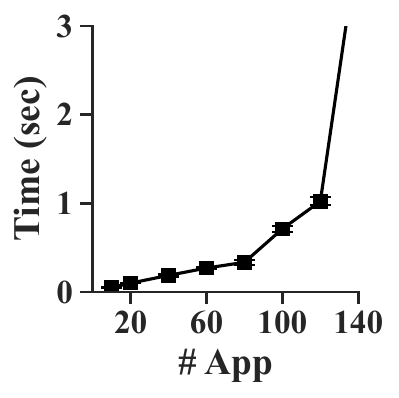}%
    \label{fig:app_time}%
    \includegraphics[width=0.24\linewidth]{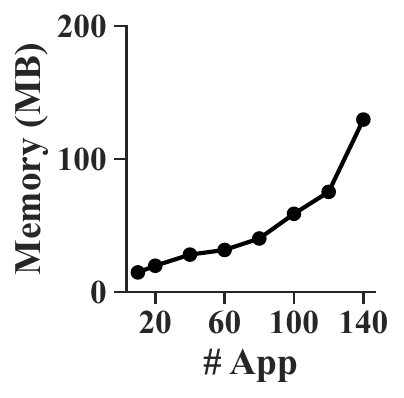}
    \label{fig:app_mem}
    }%
    \caption{Scalability of \proposedsystem to input parameters.} 
    \label{fig:9_scalability}
    \vspace{-5mm}
\end{figure}

We evaluate the runtime performance of \proposedsystem in the mesoscale regional edge deployments. When a workload arrives, \proposedsystem requires approximately 3.3 $ms$ to determine the placement and  1.01 $s$ to initiate the application deployment. We further analyze the overhead of our incremental placement algorithm, which directly affects system performance. ~\autoref{fig:9_scalability} shows how the runtime and memory usage of our algorithm scale with two key parameters: the number of servers and the number of applications. By isolating these parameters (varying one while fixing the other), we demonstrate that our incremental placement algorithm scales efficiently to 400 servers and 140 applications, completing computations within 3 seconds while consuming less than 200 MB of memory. 


\section{Discussion}
\label{sec:discussion}
We have shown the potential of mesoscale carbon intensity information in decarbonizing edge computing. In this section, we reflect on the adoption of \proposedsystem for futuristic edge infrastructures and the limitations.

\noindent {\bf Adopting \proposedsystem.}
Although the carbon intensity in many mesoscale regions is still opaque, we have shown that mesoscale spatial shifting can effectively reduce carbon emissions. In addition, the advances in carbon accounting and the continuing adoption of local renewable energy will enable fine-grained carbon information at a regional and even at the edge data center level. In this case, as highlighted in earlier research~\cite{acun2023carbon, Zheng2020:Curtailment, Gsteiger2024:Caribou}, carbon-aware spatial shifting becomes more crucial in the decarbonization of computing.

\noindent {\bf Limitations.} Currently, \proposedsystem does not automatically redeploy applications to adapt to dynamic workloads and changes in carbon intensity, ensuring low computational overhead and service continuity. Additionally, while carbon-aware spatial shifting has potential and applicability for edge applications, our evaluations were limited to edge data centers with available data from the Akamai CDN traces and the WonderNetwork traces\cite{wonder-proxy-2020}.



\noindent {\bf Holistic Emissions Reduction.}
In this paper, we only focused on the {\em operational emissions} of edge workloads from energy consumption, while the {\em embodied emissions} from manufacturing servers is beyond the scope of this paper\cite{Eeckhout2024:FOCAL, Gupta2022:ACT, Li2023:Toward, Gupta2022:Chasing}. Nonetheless, we note that \proposedsystem does not require increases in the number of servers, and earlier research has shown how spatial shifting can help extend the hardware lifespan, amortizing its embodied emissions~\cite{Liu2024:Relocation, Switzer2023:Junkyard}.






\section{Related Work}
\label{sec:related_work}


Researchers have analyzed the potential of spatial shifting for batch workloads~\cite{cloudcarbon, sukprasert2024limitations, Zheng2020:Curtailment, Lin2023:Adapting}. For instance, \cite{cloudcarbon} have shown how spatial shifting can reduce the carbon emissions of machine learning training workloads, while in \cite{Zheng2020:Curtailment}, researchers have analyzed how spatial shifting can utilize curtailed energy, which increases the utility of renewable energy.
In addition to batch workloads, researchers have shown how interactive workloads can benefit from spatial shifting~\cite{sukprasert2024limitations, igsc2023-casper, Baolin2023:Clover, Chadha2023:GreenCourier, maji_hotcarbon23, Gsteiger2024:Caribou, Gao-2012-being-green, Murillo2024:CDNShifter}. For instance, in \cite{sukprasert2024limitations, igsc2023-casper}, the authors demonstrated that spatial shifting across geographically dispersed cloud data centers can reduce the carbon emissions of web requests. Moreover,~\cite{Baolin2023:Clover} analyzed how spatial shifting can be used for machine learning inference, where the authors showed how combining spatial shifting with model selection can reduce carbon emissions further. In this paper, we underscore the potential of geospatial workload shifting across mesoscale edge data centers, where the benefits of carbon savings outweigh the cost of latency increases. Moreover, we propose a carbon-aware framework for optimizing edge application placements to reduce emissions at the edge. 
Lastly, many researchers have highlighted many prevalent trade-offs in carbon-aware optimizations to include carbon-performance trade-offs \cite{wait-awhile, igsc2023-casper}, carbon-energy trade-offs~\cite{Baolin2023:Clover, Gupta2022:Chasing, Jiang204:EcoLife}, carbon-accuracy\cite{Baolin2023:Clover}, carbon-cost trade-offs~\cite{Gao-2012-being-green, Murillo2024:CDNShifter}.
This paper considers the carbon-performance and carbon-energy trade-offs that are more prevalent in edge computing.

\section{Conclusion}
\label{sec:conclusion}
In this paper, we analyzed fine-grained carbon intensity traces at intermediate ``mesoscales,'' such as within a single U.S. state or neighboring countries in Europe, and showed that intelligently distributing workload at these mesoscales can reduce carbon without violating latency SLOs.
To build upon this observation, we presented \proposedsystem, a carbon-aware placement framework for edge applications, which optimizes workload placement and power management decisions across edge data centers within a mesoscale region to minimize carbon emissions while satisfying latency SLOs. 
Our evaluation highlights that our \proposedsystem can exploit the mesoscale carbon intensity variations and present carbon savings that outweigh the latency overhead. 
Current work does not consider the cost of data movement across edge locations. In future work, we will enhance \proposedsystem to consider the data movement cost, such as latency and the carbon emissions associated with storage and transfer.  



\begin{acks}
We thank the anonymous HPDC reviewers for their valuable insight and feedback, and Electricity Maps for access to their
carbon intensity data. This research is supported by National Science Foundation (NSF) grants 2213636, 
2105494, 2211302, 2211888, 2325956, the U.S. Department of Energy Award DE-EE0010143, and support from VMware. This work used Amazon Web Services through the CloudBank, which is supported by NSF grant 19250001. The creation of Sinfonia was supported by the NSF under grant number CNS-2106862.

\end{acks}

\bibliographystyle{ACM-Reference-Format}
\bibliography{paper}


\begin{thebibliography}{44}


\ifx \showCODEN    \undefined \def \showCODEN     #1{\unskip}     \fi
\ifx \showDOI      \undefined \def \showDOI       #1{#1}\fi
\ifx \showISBNx    \undefined \def \showISBNx     #1{\unskip}     \fi
\ifx \showISBNxiii \undefined \def \showISBNxiii  #1{\unskip}     \fi
\ifx \showISSN     \undefined \def \showISSN      #1{\unskip}     \fi
\ifx \showLCCN     \undefined \def \showLCCN      #1{\unskip}     \fi
\ifx \shownote     \undefined \def \shownote      #1{#1}          \fi
\ifx \showarticletitle \undefined \def \showarticletitle #1{#1}   \fi
\ifx \showURL      \undefined \def \showURL       {\relax}        \fi
\providecommand\bibfield[2]{#2}
\providecommand\bibinfo[2]{#2}
\providecommand\natexlab[1]{#1}
\providecommand\showeprint[2][]{arXiv:#2}

\bibitem[wat(2024)]%
        {watttime}
 \bibinfo{year}{2024}\natexlab{}.
\newblock \bibinfo{title}{Watt{T}ime}.
\newblock \bibinfo{howpublished}{\url{https://www.watttime.org/}}.
\newblock


\bibitem[Acun et~al\mbox{.}(2023)]%
        {acun2023carbon}
\bibfield{author}{\bibinfo{person}{Bilge Acun}, \bibinfo{person}{Benjamin Lee}, \bibinfo{person}{Fiodar Kazhamiaka}, \bibinfo{person}{Kiwan Maeng}, \bibinfo{person}{Udit Gupta}, \bibinfo{person}{Manoj Chakkaravarthy}, \bibinfo{person}{David Brooks}, {and} \bibinfo{person}{Carole-Jean Wu}.} \bibinfo{year}{2023}\natexlab{}.
\newblock \showarticletitle{{Carbon Explorer: A Holistic Framework for Designing Carbon Aware Datacenters}}. In \bibinfo{booktitle}{\emph{ACM International Conference on Architectural Support for Programming Languages and Operating Systems (ASPLOS)}}. \bibinfo{pages}{118--132}.
\newblock


\bibitem[Bashir et~al\mbox{.}(2022)]%
        {Bashir2022:HotAir}
\bibfield{author}{\bibinfo{person}{Noman Bashir}, \bibinfo{person}{David Irwin}, \bibinfo{person}{Prashant Shenoy}, {and} \bibinfo{person}{Abel Souza}.} \bibinfo{year}{2022}\natexlab{}.
\newblock \showarticletitle{Sustainable {C}omputing -- {W}ithout the {H}ot {A}ir}. In \bibinfo{booktitle}{\emph{Proceedings of the First Workshop on Sustainable Computer Systems Design and Implementation (HotCarbon)}}.
\newblock


\bibitem[Bochkovskiy et~al\mbox{.}(2020)]%
        {bochkovskiy2020yolov4}
\bibfield{author}{\bibinfo{person}{Alexey Bochkovskiy}, \bibinfo{person}{Chien-Yao Wang}, {and} \bibinfo{person}{Hong-Yuan~Mark Liao}.} \bibinfo{year}{2020}\natexlab{}.
\newblock \showarticletitle{{YOLOv4: Optimal Speed and Accuracy of Object Detection}}.
\newblock \bibinfo{journal}{\emph{arXiv preprint arXiv:2004.10934}} (\bibinfo{year}{2020}).
\newblock
\urldef\tempurl%
\url{https://arxiv.org/abs/2004.10934}
\showURL{%
\tempurl}


\bibitem[Cai et~al\mbox{.}(2017)]%
        {cai2017neuralpower}
\bibfield{author}{\bibinfo{person}{Ermao Cai}, \bibinfo{person}{Da-Cheng Juan}, \bibinfo{person}{Dimitrios Stamoulis}, {and} \bibinfo{person}{Diana Marculescu}.} \bibinfo{year}{2017}\natexlab{}.
\newblock \showarticletitle{Neuralpower: {P}redict and {D}eploy {E}nergy-efficient {C}onvolutional {N}eural {N}etworks}. In \bibinfo{booktitle}{\emph{Asian Conference on Machine Learning}}.
\newblock


\bibitem[Calma(2024)]%
        {google_offshore_wind_2024}
\bibfield{author}{\bibinfo{person}{Justine Calma}.} \bibinfo{year}{2024}\natexlab{}.
\newblock \showarticletitle{Google to power data centers in Europe with offshore wind farms}.
\newblock \bibinfo{journal}{\emph{The Verge}} (\bibinfo{year}{2024}).
\newblock
\urldef\tempurl%
\url{https://www.theverge.com/2024/2/1/24056819/google-offshore-wind-farms-data-centers-europe}
\showURL{%
\tempurl}
\newblock
\shownote{Accessed: 2024-02-01}.


\bibitem[Chadha et~al\mbox{.}(2023)]%
        {Chadha2023:GreenCourier}
\bibfield{author}{\bibinfo{person}{Mohak Chadha}, \bibinfo{person}{Thandayuthapani Subramanian}, \bibinfo{person}{Eishi Arima}, \bibinfo{person}{Michael Gerndt}, \bibinfo{person}{Martin Schulz}, {and} \bibinfo{person}{Osama Abboud}.} \bibinfo{year}{2023}\natexlab{}.
\newblock \showarticletitle{{GreenCourier: Carbon-Aware Scheduling for Serverless Functions}}. In \bibinfo{booktitle}{\emph{Proceedings of the 9th International Workshop on Serverless Computing}} (Bologna, Italy) \emph{(\bibinfo{series}{WoSC '23})}. \bibinfo{pages}{18–23}.
\newblock
\showISBNx{9798400704550}
\urldef\tempurl%
\url{https://doi.org/10.1145/3631295.3631396}
\showDOI{\tempurl}


\bibitem[David et~al\mbox{.}(2010)]%
        {david2010rapl}
\bibfield{author}{\bibinfo{person}{Howard David}, \bibinfo{person}{Eugene Gorbatov}, \bibinfo{person}{Ulf~R Hanebutte}, \bibinfo{person}{Rahul Khanna}, {and} \bibinfo{person}{Christian Le}.} \bibinfo{year}{2010}\natexlab{}.
\newblock \showarticletitle{R{A}{P}{L}: {M}emory {P}ower {E}stimation and {C}apping}. In \bibinfo{booktitle}{\emph{ACM/IEEE International Symposium on Low-Power Electronics and Design (ISLPED)}}.
\newblock


\bibitem[Dodge et~al\mbox{.}(2022)]%
        {cloudcarbon}
\bibfield{author}{\bibinfo{person}{Jesse Dodge}, \bibinfo{person}{Taylor Prewitt}, \bibinfo{person}{Remi Tachet~des Combes}, \bibinfo{person}{Erika Odmark}, \bibinfo{person}{Roy Schwartz}, \bibinfo{person}{Emma Strubell}, \bibinfo{person}{Alexandra~Sasha Luccioni}, \bibinfo{person}{Noah~A. Smith}, \bibinfo{person}{Nicole DeCario}, {and} \bibinfo{person}{Will Buchanan}.} \bibinfo{year}{2022}\natexlab{}.
\newblock \showarticletitle{Measuring the Carbon Intensity of AI in Cloud Instances}. In \bibinfo{booktitle}{\emph{2022 ACM Conference on Fairness, Accountability, and Transparency}} \emph{(\bibinfo{series}{FAccT '22})}. \bibinfo{pages}{1877--1894}.
\newblock
\showISBNx{9781450393522}


\bibitem[Eeckhout(2024)]%
        {Eeckhout2024:FOCAL}
\bibfield{author}{\bibinfo{person}{Lieven Eeckhout}.} \bibinfo{year}{2024}\natexlab{}.
\newblock \showarticletitle{{FOCAL: A First-Order Carbon Model to Assess Processor Sustainability}}. In \bibinfo{booktitle}{\emph{Proceedings of the 29th ACM International Conference on Architectural Support for Programming Languages and Operating Systems, Volume 2}} (La Jolla, CA, USA) \emph{(\bibinfo{series}{ASPLOS '24})}. \bibinfo{pages}{401–415}.
\newblock
\showISBNx{9798400703850}
\urldef\tempurl%
\url{https://doi.org/10.1145/3620665.3640415}
\showDOI{\tempurl}


\bibitem[Gao et~al\mbox{.}(2012)]%
        {Gao-2012-being-green}
\bibfield{author}{\bibinfo{person}{Peter~Xiang Gao}, \bibinfo{person}{Andrew~R. Curtis}, \bibinfo{person}{Bernard Wong}, {and} \bibinfo{person}{Srinivasan Keshav}.} \bibinfo{year}{2012}\natexlab{}.
\newblock \showarticletitle{{It’s Not Easy Being Green}}. In \bibinfo{booktitle}{\emph{Proceedings of the ACM SIGCOMM 2012 Conference on Applications, Technologies, Architectures, and Protocols for Computer Communication}} (Helsinki, Finland) \emph{(\bibinfo{series}{SIGCOMM '12})}. \bibinfo{pages}{211–222}.
\newblock
\showISBNx{9781450314190}
\urldef\tempurl%
\url{https://doi.org/10.1145/2342356.2342398}
\showDOI{\tempurl}


\bibitem[Goudarzi et~al\mbox{.}(2021)]%
        {Goudarzi21Placement}
\bibfield{author}{\bibinfo{person}{Mohammad Goudarzi}, \bibinfo{person}{Huaming Wu}, \bibinfo{person}{Marimuthu Palaniswami}, {and} \bibinfo{person}{Rajkumar Buyya}.} \bibinfo{year}{2021}\natexlab{}.
\newblock \showarticletitle{{An Application Placement Technique for Concurrent IoT Applications in Edge and Fog Computing Environments}}.
\newblock \bibinfo{journal}{\emph{IEEE Transactions on Mobile Computing}} \bibinfo{volume}{20}, \bibinfo{number}{4} (\bibinfo{year}{2021}), \bibinfo{pages}{1298--1311}.
\newblock
\urldef\tempurl%
\url{https://doi.org/10.1109/TMC.2020.2967041}
\showDOI{\tempurl}


\bibitem[Gsteiger et~al\mbox{.}(2024)]%
        {Gsteiger2024:Caribou}
\bibfield{author}{\bibinfo{person}{Viktor~Urban Gsteiger}, \bibinfo{person}{Pin Hong~(Daniel) Long}, \bibinfo{person}{Yiran~(Jerry) Sun}, \bibinfo{person}{Parshan Javanrood}, {and} \bibinfo{person}{Mohammad Shahrad}.} \bibinfo{year}{2024}\natexlab{}.
\newblock \showarticletitle{{Caribou: Fine-Grained Geospatial Shifting of Serverless Applications for Sustainability}}. In \bibinfo{booktitle}{\emph{Proceedings of the ACM SIGOPS 30th Symposium on Operating Systems Principles}} (Austin, TX, USA) \emph{(\bibinfo{series}{SOSP '24})}. \bibinfo{pages}{403–420}.
\newblock
\showISBNx{9798400712517}
\urldef\tempurl%
\url{https://doi.org/10.1145/3694715.3695954}
\showDOI{\tempurl}


\bibitem[Gupta et~al\mbox{.}(2022a)]%
        {Gupta2022:ACT}
\bibfield{author}{\bibinfo{person}{Udit Gupta}, \bibinfo{person}{Mariam Elgamal}, \bibinfo{person}{Gage Hills}, \bibinfo{person}{Gu-Yeon Wei}, \bibinfo{person}{Hsien-Hsin~S. Lee}, \bibinfo{person}{David Brooks}, {and} \bibinfo{person}{Carole-Jean Wu}.} \bibinfo{year}{2022}\natexlab{a}.
\newblock \showarticletitle{{ACT: Designing Sustainable Computer Systems With An Architectural Carbon Modeling Tool}}. In \bibinfo{booktitle}{\emph{Proceedings of the 49th Annual International Symposium on Computer Architecture}} (New York, New York) \emph{(\bibinfo{series}{ISCA '22})}. \bibinfo{pages}{784–799}.
\newblock
\showISBNx{9781450386104}
\urldef\tempurl%
\url{https://doi.org/10.1145/3470496.3527408}
\showDOI{\tempurl}


\bibitem[Gupta et~al\mbox{.}(2022b)]%
        {Gupta2022:Chasing}
\bibfield{author}{\bibinfo{person}{Udit Gupta}, \bibinfo{person}{Young~Geun Kim}, \bibinfo{person}{Sylvia Lee}, \bibinfo{person}{Jordan Tse}, \bibinfo{person}{Hsien-Hsin~S. Lee}, \bibinfo{person}{Gu-Yeon Wei}, \bibinfo{person}{David Brooks}, {and} \bibinfo{person}{Carole-Jean Wu}.} \bibinfo{year}{2022}\natexlab{b}.
\newblock \showarticletitle{{Chasing Carbon: The Elusive Environmental Footprint of Computing}}.
\newblock \bibinfo{journal}{\emph{IEEE Micro}} \bibinfo{volume}{42}, \bibinfo{number}{4} (\bibinfo{date}{jul} \bibinfo{year}{2022}), \bibinfo{pages}{37–47}.
\newblock
\showISSN{0272-1732}
\urldef\tempurl%
\url{https://doi.org/10.1109/MM.2022.3163226}
\showDOI{\tempurl}


\bibitem[Hanafy et~al\mbox{.}(2023)]%
        {hanafy2023carbonscaler}
\bibfield{author}{\bibinfo{person}{Walid~A. Hanafy}, \bibinfo{person}{Qianlin Liang}, \bibinfo{person}{Noman Bashir}, \bibinfo{person}{David Irwin}, {and} \bibinfo{person}{Prashant Shenoy}.} \bibinfo{year}{2023}\natexlab{}.
\newblock \showarticletitle{{CarbonScaler: Leveraging Cloud Workload Elasticity for Optimizing Carbon-Efficiency}}.
\newblock \bibinfo{journal}{\emph{Proceedings of the ACM on Measurement and Analysis of Computing Systems}} \bibinfo{volume}{7}, \bibinfo{number}{3}, Article \bibinfo{articleno}{57} (\bibinfo{date}{December} \bibinfo{year}{2023}), \bibinfo{numpages}{28}~pages.
\newblock
\urldef\tempurl%
\url{https://doi.org/10.1145/3626788}
\showDOI{\tempurl}


\bibitem[He et~al\mbox{.}(2016)]%
        {resnet}
\bibfield{author}{\bibinfo{person}{Kaiming He}, \bibinfo{person}{Xiangyu Zhang}, \bibinfo{person}{Shaoqing Ren}, {and} \bibinfo{person}{Jian Sun}.} \bibinfo{year}{2016}\natexlab{}.
\newblock \showarticletitle{Deep {R}esidual {L}earning for {I}mage {R}ecognition}. In \bibinfo{booktitle}{\emph{Proceedings of the IEEE conference on computer vision and pattern recognition (CVPR)}}.
\newblock


\bibitem[{International Energy Agency}(2024)]%
        {iea2024electricity}
\bibfield{author}{\bibinfo{person}{{International Energy Agency}}.} \bibinfo{year}{2024}\natexlab{}.
\newblock \bibinfo{booktitle}{\emph{Electricity 2024}}.
\newblock \bibinfo{type}{{T}echnical {R}eport}. \bibinfo{institution}{IEA}, \bibinfo{address}{Paris}.
\newblock
\urldef\tempurl%
\url{https://www.iea.org/reports/electricity-2024}
\showURL{%
\tempurl}


\bibitem[Jiang et~al\mbox{.}(2024)]%
        {Jiang204:EcoLife}
\bibfield{author}{\bibinfo{person}{Yankai Jiang}, \bibinfo{person}{Rohan~Basu Roy}, \bibinfo{person}{Baolin Li}, {and} \bibinfo{person}{Devesh Tiwari}.} \bibinfo{year}{2024}\natexlab{}.
\newblock \showarticletitle{{EcoLife: Carbon-Aware Serverless Function Scheduling for Sustainable Computing}}. In \bibinfo{booktitle}{\emph{Proceedings of the International Conference for High Performance Computing, Networking, Storage, and Analysis}} (Atlanta, GA, USA) \emph{(\bibinfo{series}{SC '24})}. \bibinfo{publisher}{IEEE Press}, Article \bibinfo{articleno}{12}, \bibinfo{numpages}{15}~pages.
\newblock
\showISBNx{9798350352917}
\urldef\tempurl%
\url{https://doi.org/10.1109/SC41406.2024.00018}
\showDOI{\tempurl}


\bibitem[Li et~al\mbox{.}(2023a)]%
        {Li2023:Toward}
\bibfield{author}{\bibinfo{person}{Baolin Li}, \bibinfo{person}{Rohan Basu~Roy}, \bibinfo{person}{Daniel Wang}, \bibinfo{person}{Siddharth Samsi}, \bibinfo{person}{Vijay Gadepally}, {and} \bibinfo{person}{Devesh Tiwari}.} \bibinfo{year}{{2023}}\natexlab{a}.
\newblock \showarticletitle{Toward Sustainable HPC: Carbon Footprint Estimation and Environmental Implications of HPC Systems}. In \bibinfo{booktitle}{\emph{Proceedings of the International Conference for High Performance Computing, Networking, Storage and Analysis}} (Denver, CO, USA) \emph{(\bibinfo{series}{SC '23})}. Article \bibinfo{articleno}{19}, \bibinfo{numpages}{15}~pages.
\newblock
\showISBNx{9798400701092}
\urldef\tempurl%
\url{https://doi.org/10.1145/3581784.3607035}
\showDOI{\tempurl}


\bibitem[Li et~al\mbox{.}(2023b)]%
        {Baolin2023:Clover}
\bibfield{author}{\bibinfo{person}{Baolin Li}, \bibinfo{person}{Siddharth Samsi}, \bibinfo{person}{Vijay Gadepally}, {and} \bibinfo{person}{Devesh Tiwari}.} \bibinfo{year}{2023}\natexlab{b}.
\newblock \showarticletitle{{Clover: Toward Sustainable AI with Carbon-Aware Machine Learning Inference Service}}. In \bibinfo{booktitle}{\emph{Proceedings of the International Conference for High Performance Computing, Networking, Storage and Analysis}} (Denver, CO, USA) \emph{(\bibinfo{series}{SC '23})}. Article \bibinfo{articleno}{20}, \bibinfo{numpages}{15}~pages.
\newblock
\showISBNx{9798400701092}
\urldef\tempurl%
\url{https://doi.org/10.1145/3581784.3607034}
\showDOI{\tempurl}


\bibitem[Li and Wang(2018)]%
        {Li2018_energy_placement}
\bibfield{author}{\bibinfo{person}{Yuanzhe Li} {and} \bibinfo{person}{Shangguang Wang}.} \bibinfo{year}{2018}\natexlab{}.
\newblock \showarticletitle{{An Energy-Aware Edge Server Placement Algorithm in Mobile Edge Computing}}. In \bibinfo{booktitle}{\emph{2018 IEEE International Conference on Edge Computing (EDGE)}}. \bibinfo{pages}{66--73}.
\newblock
\urldef\tempurl%
\url{https://doi.org/10.1109/EDGE.2018.00016}
\showDOI{\tempurl}


\bibitem[Lin and Chien(2023)]%
        {Lin2023:Adapting}
\bibfield{author}{\bibinfo{person}{Liuzixuan Lin} {and} \bibinfo{person}{Andrew~A. Chien}.} \bibinfo{year}{2023}\natexlab{}.
\newblock \showarticletitle{{Adapting Datacenter Capacity for Greener Datacenters and Grid}}. In \bibinfo{booktitle}{\emph{{Proceedings of the 14th ACM International Conference on Future Energy Systems}}} (Orlando, FL, USA) \emph{(\bibinfo{series}{e-Energy '23})}. \bibinfo{pages}{200–213}.
\newblock
\showISBNx{9798400700323}
\urldef\tempurl%
\url{https://doi.org/10.1145/3575813.3595197}
\showDOI{\tempurl}


\bibitem[Liu et~al\mbox{.}(2024)]%
        {Liu2024:Relocation}
\bibfield{author}{\bibinfo{person}{Yejia Liu}, \bibinfo{person}{Pengfei Li}, \bibinfo{person}{Daniel Wong}, {and} \bibinfo{person}{Shaolei Ren}.} \bibinfo{year}{2024}\natexlab{}.
\newblock \showarticletitle{{Geographical Server Relocation: Opportunities and Challenges}}. In \bibinfo{booktitle}{\emph{Proceedings of HotCarbon'24}}.
\newblock
\urldef\tempurl%
\url{https://hotcarbon.org/assets/2024/pdf/hotcarbon24-final20.pdf}
\showURL{%
\tempurl}
\newblock
\shownote{Accessed: 2024-10-08}.


\bibitem[Maji et~al\mbox{.}(2023)]%
        {maji_hotcarbon23}
\bibfield{author}{\bibinfo{person}{Diptyaroop Maji}, \bibinfo{person}{Ben Pfaff}, \bibinfo{person}{Vipin~P. R}, \bibinfo{person}{Rajagopal Sreenivasan}, \bibinfo{person}{Victor Firoiu}, \bibinfo{person}{Sreeram Iyer}, \bibinfo{person}{Colleen Josephson}, \bibinfo{person}{Zhelong Pan}, {and} \bibinfo{person}{Ramesh~K. Sitaraman}.} \bibinfo{year}{2023}\natexlab{}.
\newblock \showarticletitle{{Bringing Carbon Awareness to Multi-cloud Application Delivery}}. In \bibinfo{booktitle}{\emph{Proceedings of the 2nd Workshop on Sustainable Computer Systems, HotCarbon 2023, Boston, MA, USA, 9 July 2023}}. \bibinfo{pages}{6:1--6:6}.
\newblock


\bibitem[Maps(2024)]%
        {electricity-map}
\bibfield{author}{\bibinfo{person}{Electricity Maps}.} \bibinfo{year}{2024}\natexlab{}.
\newblock \bibinfo{title}{Electricity {M}aps}.
\newblock \bibinfo{howpublished}{\url{https://www.electricitymap.org/map}}.
\newblock


\bibitem[Murillo et~al\mbox{.}(2024)]%
        {Murillo2024:CDNShifter}
\bibfield{author}{\bibinfo{person}{Jorge Murillo}, \bibinfo{person}{Walid~A. Hanafy}, \bibinfo{person}{David Irwin}, \bibinfo{person}{Ramesh Sitaraman}, {and} \bibinfo{person}{Prashant Shenoy}.} \bibinfo{year}{2024}\natexlab{}.
\newblock \showarticletitle{{CDN-Shifter: Leveraging Spatial Workload Shifting to Decarbonize Content Delivery Networks}}. In \bibinfo{booktitle}{\emph{Proceedings of the 2024 ACM Symposium on Cloud Computing}} (Redmond, WA, USA) \emph{(\bibinfo{series}{SoCC '24})}. \bibinfo{publisher}{Association for Computing Machinery}, \bibinfo{address}{New York, NY, USA}, \bibinfo{pages}{505–521}.
\newblock
\showISBNx{9798400712869}
\urldef\tempurl%
\url{https://doi.org/10.1145/3698038.3698516}
\showDOI{\tempurl}


\bibitem[{NVIDIA}(2023)]%
        {nvidia_dcgm_exporter_github}
\bibfield{author}{\bibinfo{person}{{NVIDIA}}.} \bibinfo{year}{2023}\natexlab{}.
\newblock \bibinfo{title}{DCGM Exporter}.
\newblock \bibinfo{howpublished}{\url{https://github.com/NVIDIA/dcgm-exporter}}.
\newblock


\bibitem[Perron and Furnon(2024)]%
        {ortools}
\bibfield{author}{\bibinfo{person}{Laurent Perron} {and} \bibinfo{person}{Vincent Furnon}.} \bibinfo{year}{2024}\natexlab{}.
\newblock \bibinfo{booktitle}{\emph{{Google OR-Tools v9.10}}}.
\newblock Google.
\newblock
\urldef\tempurl%
\url{https://developers.google.com/optimization/}
\showURL{%
\tempurl}


\bibitem[Qi et~al\mbox{.}(2017)]%
        {paleo}
\bibfield{author}{\bibinfo{person}{Qi}, \bibinfo{person}{Evan~R. Sparks}, {and} \bibinfo{person}{Ameet~S. Talwalkar}.} \bibinfo{year}{2017}\natexlab{}.
\newblock \showarticletitle{Paleo: A Performance Model for Deep Neural Networks}. In \bibinfo{booktitle}{\emph{The International Conference on Learning Representations}} \emph{(\bibinfo{series}{ICLR'17})}.
\newblock


\bibitem[Rabenstein and Volz(2015)]%
        {Prometheus}
\bibfield{author}{\bibinfo{person}{Bj{\"o}rn Rabenstein} {and} \bibinfo{person}{Julius Volz}.} \bibinfo{year}{2015}\natexlab{}.
\newblock \showarticletitle{Prometheus: A {Next-Generation} Monitoring System (Talk)}. \bibinfo{publisher}{USENIX Association}, \bibinfo{address}{Dublin}.
\newblock


\bibitem[Reinheimer(2020)]%
        {wonder-proxy-2020}
\bibfield{author}{\bibinfo{person}{Paul Reinheimer}.} \bibinfo{year}{2020}\natexlab{}.
\newblock \bibinfo{title}{{A day in the life of the Internet}}.
\newblock \bibinfo{howpublished}{\url{https://wonderproxy.com/blog/a-day-in-the-life-of-the-internet/}}.
\newblock
\newblock
\shownote{Accessed: (2023-04-14)}.


\bibitem[Satyanarayanan(2017)]%
        {Satya17_emergence}
\bibfield{author}{\bibinfo{person}{Mahadev Satyanarayanan}.} \bibinfo{year}{2017}\natexlab{}.
\newblock \showarticletitle{{The Emergence of Edge Computing}}.
\newblock \bibinfo{journal}{\emph{Computer}} \bibinfo{volume}{50}, \bibinfo{number}{1} (\bibinfo{year}{2017}), \bibinfo{pages}{30--39}.
\newblock
\urldef\tempurl%
\url{https://doi.org/10.1109/MC.2017.9}
\showDOI{\tempurl}


\bibitem[Satyanarayanan et~al\mbox{.}(2009)]%
        {Satya09_Cloudlets}
\bibfield{author}{\bibinfo{person}{Mahadev Satyanarayanan}, \bibinfo{person}{Paramvir Bahl}, \bibinfo{person}{Ramon Caceres}, {and} \bibinfo{person}{Nigel Davies}.} \bibinfo{year}{2009}\natexlab{}.
\newblock \showarticletitle{{The Case for VM-Based Cloudlets in Mobile Computing}}.
\newblock \bibinfo{journal}{\emph{IEEE Pervasive Computing}} \bibinfo{volume}{8}, \bibinfo{number}{4} (\bibinfo{year}{2009}), \bibinfo{pages}{14--23}.
\newblock
\urldef\tempurl%
\url{https://doi.org/10.1109/MPRV.2009.82}
\showDOI{\tempurl}


\bibitem[Satyanarayanan et~al\mbox{.}(2022)]%
        {satyanarayanan2022sinfonia}
\bibfield{author}{\bibinfo{person}{Mahadev Satyanarayanan}, \bibinfo{person}{Jan Harkes}, \bibinfo{person}{Jim Blakley}, \bibinfo{person}{Marc Meunier}, \bibinfo{person}{Govindarajan Mohandoss}, \bibinfo{person}{Kiel Friedt}, \bibinfo{person}{Arun Thulasi}, \bibinfo{person}{Pranav Saxena}, {and} \bibinfo{person}{Brian Barritt}.} \bibinfo{year}{2022}\natexlab{}.
\newblock \showarticletitle{{Sinfonia: Cross-Tier Orchestration for Edge-Native Applications}}.
\newblock \bibinfo{journal}{\emph{Frontiers in the Internet of Things}}  \bibinfo{volume}{1} (\bibinfo{year}{2022}), \bibinfo{pages}{1025247}.
\newblock


\bibitem[Souza et~al\mbox{.}(2023a)]%
        {ecovisor}
\bibfield{author}{\bibinfo{person}{Abel Souza}, \bibinfo{person}{Noman Bashir}, \bibinfo{person}{Jorge Murillo}, \bibinfo{person}{Walid Hanafy}, \bibinfo{person}{Qianlin Liang}, \bibinfo{person}{David Irwin}, {and} \bibinfo{person}{Prashant Shenoy}.} \bibinfo{year}{2023}\natexlab{a}.
\newblock \showarticletitle{Ecovisor: A {V}irtual {E}nergy {S}ystem for {C}arbon-{E}fficient {A}pplications}. In \bibinfo{booktitle}{\emph{ACM International Conference on Architectural Support for Programming Languages and Operating Systems (ASPLOS)}}. \bibinfo{pages}{252--265}.
\newblock


\bibitem[Souza et~al\mbox{.}(2023b)]%
        {igsc2023-casper}
\bibfield{author}{\bibinfo{person}{Abel Souza}, \bibinfo{person}{Shruti Jasoria}, \bibinfo{person}{Basundhara Chakrabarty}, \bibinfo{person}{Alexander Bridgwater}, \bibinfo{person}{Axel Lundberg}, \bibinfo{person}{Filip Skogh}, \bibinfo{person}{Ahmed Ali-Eldin}, \bibinfo{person}{David Irwin}, {and} \bibinfo{person}{Prashant Shenoy}.} \bibinfo{year}{2023}\natexlab{b}.
\newblock \showarticletitle{{CASPER: Carbon-Aware Scheduling and Provisioning for Distributed Web Services}}. In \bibinfo{booktitle}{\emph{Proceedings of the 14th International Green and Sustainable Computing Conference (IGSC), Toronto, ON, Canada}}.
\newblock


\bibitem[Sukprasert et~al\mbox{.}(2024)]%
        {sukprasert2024limitations}
\bibfield{author}{\bibinfo{person}{Thanathorn Sukprasert}, \bibinfo{person}{Abel Souza}, \bibinfo{person}{Noman Bashir}, \bibinfo{person}{David Irwin}, {and} \bibinfo{person}{Prashant Shenoy}.} \bibinfo{year}{2024}\natexlab{}.
\newblock \showarticletitle{{On the Limitations of Carbon-Aware Temporal and Spatial Workload Shifting in the Cloud}}. In \bibinfo{booktitle}{\emph{Proceedings of the Nineteenth European Conference on Computer Systems}} \emph{(\bibinfo{series}{EuroSys '24})}. \bibinfo{pages}{924–941}.
\newblock
\showISBNx{9798400704376}
\urldef\tempurl%
\url{https://doi.org/10.1145/3627703.3650079}
\showDOI{\tempurl}


\bibitem[Switzer et~al\mbox{.}(2023)]%
        {Switzer2023:Junkyard}
\bibfield{author}{\bibinfo{person}{Jennifer Switzer}, \bibinfo{person}{Gabriel Marcano}, \bibinfo{person}{Ryan Kastner}, {and} \bibinfo{person}{Pat Pannuto}.} \bibinfo{year}{2023}\natexlab{}.
\newblock \showarticletitle{Junkyard Computing: Repurposing Discarded Smartphones to Minimize Carbon}. In \bibinfo{booktitle}{\emph{Proceedings of the 28th ACM International Conference on Architectural Support for Programming Languages and Operating Systems, Volume 2}} (Vancouver, BC, Canada) \emph{(\bibinfo{series}{ASPLOS 2023})}. \bibinfo{pages}{400–412}.
\newblock
\showISBNx{9781450399166}
\urldef\tempurl%
\url{https://doi.org/10.1145/3575693.3575710}
\showDOI{\tempurl}


\bibitem[Tan and Le(2019)]%
        {efficient-net}
\bibfield{author}{\bibinfo{person}{Mingxing Tan} {and} \bibinfo{person}{Quoc Le}.} \bibinfo{year}{2019}\natexlab{}.
\newblock \showarticletitle{{EfficientNet: Rethinking Model Scaling for Convolutional Neural Networks}}. In \bibinfo{booktitle}{\emph{Proceedings of the 36th International Conference on Machine Learning}} \emph{(\bibinfo{series}{Proceedings of Machine Learning Research}, Vol.~\bibinfo{volume}{97})}. \bibinfo{publisher}{PMLR}, \bibinfo{pages}{6105--6114}.
\newblock
\urldef\tempurl%
\url{https://proceedings.mlr.press/v97/tan19a.html}
\showURL{%
\tempurl}


\bibitem[Wiesner et~al\mbox{.}(2021)]%
        {wait-awhile}
\bibfield{author}{\bibinfo{person}{Philipp Wiesner}, \bibinfo{person}{Ilja Behnke}, \bibinfo{person}{Dominik Scheinert}, \bibinfo{person}{Kordian Gontarska}, {and} \bibinfo{person}{Lauritz Thamsen}.} \bibinfo{year}{2021}\natexlab{}.
\newblock \showarticletitle{Let's {W}ait {A}while: How {T}emporal {W}orkload {S}hifting {C}an {R}educe {C}arbon {E}missions in the {C}loud}. In \bibinfo{booktitle}{\emph{Proceedings of the 22nd International Middleware Conference (Middleware)}}. \bibinfo{pages}{260–272}.
\newblock


\bibitem[Yi et~al\mbox{.}(2017)]%
        {yi2017:Lavea}
\bibfield{author}{\bibinfo{person}{Shanhe Yi}, \bibinfo{person}{Zijiang Hao}, \bibinfo{person}{Qingyang Zhang}, \bibinfo{person}{Quan Zhang}, \bibinfo{person}{Weisong Shi}, {and} \bibinfo{person}{Qun Li}.} \bibinfo{year}{2017}\natexlab{}.
\newblock \showarticletitle{Lavea: Latency-aware video analytics on edge computing platform}. In \bibinfo{booktitle}{\emph{Proceedings of the Second ACM/IEEE Symposium on Edge Computing}}. \bibinfo{pages}{1--13}.
\newblock


\bibitem[Zhang et~al\mbox{.}(2019)]%
        {HeteroEdge}
\bibfield{author}{\bibinfo{person}{Daniel~(Yue) Zhang}, \bibinfo{person}{Tahmid Rashid}, \bibinfo{person}{Xukun Li}, \bibinfo{person}{Nathan Vance}, {and} \bibinfo{person}{Dong Wang}.} \bibinfo{year}{2019}\natexlab{}.
\newblock \showarticletitle{{HeteroEdge: taming the heterogeneity of edge computing system in social sensing}}. In \bibinfo{booktitle}{\emph{Proceedings of the International Conference on Internet of Things Design and Implementation}} (Montreal, Quebec, Canada) \emph{(\bibinfo{series}{IoTDI '19})}. \bibinfo{pages}{37–48}.
\newblock
\showISBNx{9781450362832}
\urldef\tempurl%
\url{https://doi.org/10.1145/3302505.3310067}
\showDOI{\tempurl}


\bibitem[Zheng et~al\mbox{.}(2020)]%
        {Zheng2020:Curtailment}
\bibfield{author}{\bibinfo{person}{Jiajia Zheng}, \bibinfo{person}{Andrew~A. Chien}, {and} \bibinfo{person}{Sangwon Suh}.} \bibinfo{year}{2020}\natexlab{}.
\newblock \showarticletitle{{Mitigating Curtailment and Carbon Emissions through Load Migration between Data Centers}}.
\newblock \bibinfo{journal}{\emph{Joule}} \bibinfo{volume}{4}, \bibinfo{number}{10} (\bibinfo{year}{2020}), \bibinfo{pages}{2208--2222}.
\newblock
\showISSN{2542-4351}
\urldef\tempurl%
\url{https://doi.org/10.1016/j.joule.2020.08.001}
\showDOI{\tempurl}


\end{thebibliography}


\end{document}